\documentclass[12pt]{article}
\usepackage{graphicx}
\usepackage{enumerate}
\usepackage{authblk}
\usepackage[utf8]{inputenc}
\usepackage[fleqn]{amsmath}
\usepackage{amsthm}
\usepackage{amssymb}
\usepackage{graphicx}
\usepackage{ragged2e}
\usepackage[font=scriptsize]{caption}
\usepackage{algorithm}
\usepackage{algpseudocode}
\usepackage{algorithm}
\usepackage{algpseudocode}
\usepackage{chngcntr}
\usepackage{mathtools}
\usepackage[english]{babel}
\usepackage{caption}
\usepackage{float}
\usepackage{subfig}
\usepackage{tabularx}
\newcounter{mycount}
\counterwithin{algorithm}{mycount}
\refstepcounter{mycount}

\newtheorem{remark}{Remark}
\newtheorem{corollary}{Corollary}
\newtheorem{definition}{Definition}
\newtheorem{theorem}{Theorem}
\newtheorem{lemma}{Lemma}
\newtheorem{assumption}{Assumption}
\newtheorem{proposition}{Proposition}

\usepackage[mathscr]{euscript}
\usepackage{url} 
\date{\today}

\begin{document}



\def\spacingset#1{\renewcommand{\baselinestretch}%
{#1}\small\normalsize} \spacingset{1}


\title{Consistent response prediction for multilayer networks on unknown manifolds}

\author[1]{Aranyak Acharyya\thanks{aachary6@jh.edu}}
\author[2]{Jesús Arroyo Relión\thanks{jarroyo@tamu.edu}}
\author[3]{Michael Clayton\thanks{mclayton@mrc-lmb.cam.ac.uk}}
\author[3]{Marta Zlatic\thanks{mzlatic@mrc-lmb.cam.ac.uk}}
\author[4]{Youngser Park\thanks{youngser@jhu.edu}}
\author[1]{Carey E. Priebe\thanks{Corresponding author:cep@jhu.edu}}

\affil[1]{Department of Applied Mathematics and Statistics, Johns Hopkins University}
\affil[2]{Department of Statistics, Texas A\&M University}
\affil[3]{MRC Laboratory of Molecular Biology, University of Cambridge}
\affil[4]{Center for Imaging Science, Johns Hopkins University}

  \maketitle

\bigskip
\begin{abstract}
Our paper deals with a collection of networks on a common set of nodes, where some of the networks are associated with responses. Assuming that the networks correspond to points on a one-dimensional manifold in a higher dimensional ambient space, we propose an algorithm to consistently predict the response at an unlabeled network. Our model involves a specific multiple random network model, namely the \textit{common subspace independent edge} model, where the networks share a common invariant subspace, and the heterogeneity amongst the networks is captured by a set of low dimensional matrices. Our algorithm estimates these low dimensional matrices that capture the heterogeneity of the networks, learns the underlying manifold by \textit{isomap}, and consistently predicts the response at an unlabeled network.  
We provide theoretical justifications for the use of our algorithm, validated by numerical simulations. Finally, we demonstrate the use of our algorithm on larval \textit{Drosophila} connectome data. 
\end{abstract}

\noindent%
\vfill

\newpage
\spacingset{1.9} 
\section{Introduction}
\label{Sec:Intro}
\begin{figure}
    \centering    \includegraphics[scale=0.80]{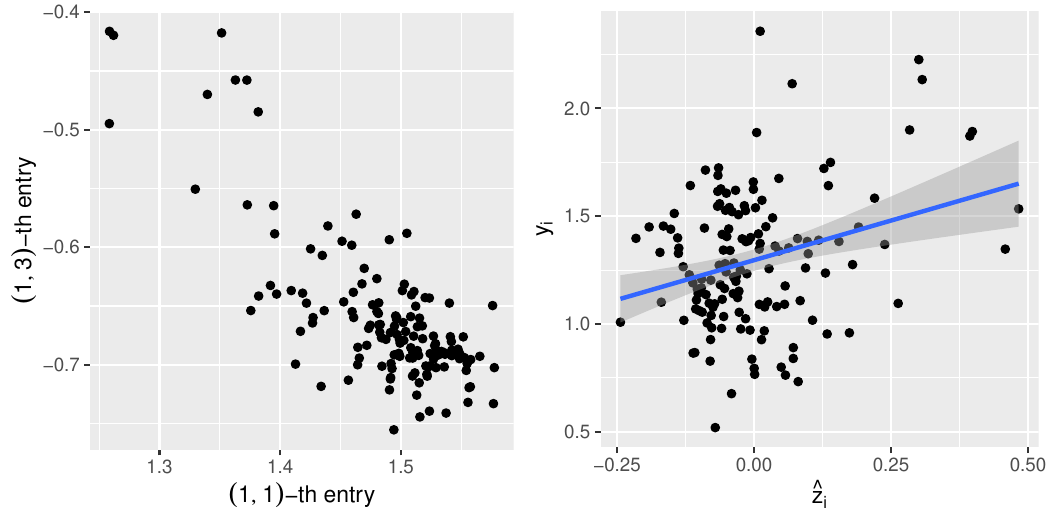}
    \caption{Application of our response prediction method to connectome data from \textit{Drosophila}. A set of networks, upon transformation through censoring, is treated as a sample from the \textit{common subspace independent edge} model. Estimates of scaled \textit{score matrices}, representing heterogeneity in the model, are obtained by multiple adjacency spectral embedding. The scatterplots and correlation coefficients obtained for all pairs of entries of the estimates of the scaled score matrices (the pair
    ($(1,1)$-th entry and $(1,3)$-th entry
     admits the maximal degree of correlation $0.780$) indicate an underlying manifold structure. The six independent components of the estimated sclaed score matrices are subsequently concatenated into six-component vectors, and \textit{isomap} is used to learn the underlying manifold. The scatterplot between the $(1,1)$-th entries and the $(1,3)$-th entries, the pair that admits the maximal degree of correlation (~$0.780$), is given on the left-hand side.
    The scatterplot between the responses and the one-dimensional isomap embeddings, along with a fitted regression line, is given on the right-hand side. An $F$-test for checking the usefulness of simple linear regression model yields a $p$-value of $0.0004$, suggesting that a simple linear regression model captures the relationship between the responses and the scalar pre-images of the points on the manifold.}
    \label{fig:intro_plot}
\end{figure}

The discipline of studying random networks has been of importance to various fields like neuroscience (\cite{Vogelstein2011GraphCU}), biology and social studies (\cite{Holland1983StochasticBF}) for a long time. Stochastic blockmodels (\cite{Holland1983StochasticBF}), where each node is assigned membership to a community and the chance of formation of an edge between two nodes depends only on their community memberships, form a popular generative model for random networks. Random dot product graphs (\cite{Young2007RandomDP}, \cite{athreya2017statistical}) are a generalization to stochastic blockmodels, where each node is assigned a feature vector also known as the latent position, and the probability of formation of an edge between two nodes equals the inner product of the corresponding latent positions. Generalized random dot product graphs (\cite{RubinDelanchy2017ASI}) are further generalization to the random dot product graphs, where the the inner product between latent positions of two nodes is replaced with indefinite inner product, to determine the probability of formation of an edge between the nodes. 
\newline
\newline
While the majority of focus in this area has been on deriving results in settings with single graphs (\cite{RubinDelanchy2017ASI}), recently, scientists have also started to study the setting of multiple graphs (\cite{Boccaletti2014TheSA},\cite{JMLR:v22:19-558},\cite{8889404}). In \cite{JMLR:v22:19-558}, the authors propose a generative model of multiple graphs with a common invariant subspace. In \cite{8889404}, an embedding method is proposed for feature extraction in multiple graphs. Works in \cite{8215766} propose a method of embedding multiple  random dot product graphs and establish a central limit theorem for the embeddings. A particular model of multiple graphs is proposed in \cite{jones2020multilayer}. In \cite{bhattacharyya2018spectral}, a spectral clustering method is proposed for community detection in multiple sparse stochastic blockmodels. Novel methods for network dimensionality reduction are proposed in \cite{omelchenko2022reducing} and \cite{Almagro2021DetectingTU}.
\newline
\newline
In this article, we consider a model of multiple graphs with a common invariant subspace, where the heterogeneity amongst the graphs is explained by a set of low-dimensional symmetric matrices (\cite{JMLR:v22:19-558}). We additionally assume that the scaled versions of these low-dimensional matrices correspond to points on a further lower-dimensional manifold. In a setting where some of the graphs are associated with scalar responses, we propose a method which exploits the presence of the underlying lower-dimensional structure to predict the response at an out-of-sample network via a linear regression model (\cite{Montgomery}). 
\newline
\newline
We illustrate the application of our theoretical results to real data (\cite{winding2023connectome},
\cite{eschbach2020recurrent}). A dataset of networks of larval \textit{Drosophila} connectome, associated with responses, is analyzed. Upon careful inspection, the presence of an underlying low-dimensional manifold structure embedded in higher dimensional ambient space is detected amongst the networks. To be more specific, we treat the collection of networks to be a sample from a particular multiple graph model (the \textit{common subspace independent edge graph model}, details in \textit{Section \ref{Sec:Definitions_and_notations}} and \cite{JMLR:v22:19-558})
and obtain low-dimensional matrices to represent the heterogeneity amongst the graphs. We compute the correlation coefficient between all pairs of entries of these matrices and observing that the highest degree of correlation between a pair indicates a strong relationship, insinuating the presence of an underlying manifold structure, we learn the manifold by applying isomap (for details see \textit{Section \ref{Subsec:Isomap}}). A scatterplot between the responses and the one-dimensional isomap embeddings indicate that a simple linear regression model can be used to explain their relationship, and an $F$-test confirms that. In \textit{Figure \ref{fig:intro_plot}}, we present the scatterplot of the components of the score matrices that exhibit the highest degree of correlation, along with the scatterplot of the responses against the one-dimensional isomap embeddings with a fitted regression line.
We use our theoretical results to establish that a simple linear regression model can be used to capture the relationship between the responses and the pre-images of the points on the manifold. 
\newline
\newline
We introduce the relevant notions and notations in \textit{Section \ref{Sec:Definitions_and_notations}}. We state the preliminaries regarding our model and multilayer stochastic blockmodel in \textit{Section \ref{Subsec:Prelims_SBM_COSIE}}, and give a brief introduction to the manifold learning technique isomap (\cite{article}) in \textit{Section \ref{Subsec:Isomap}}. In \textit{Section \ref{Sec:Model_and_Method}} we give an elaborate introduction to our model and then we formally state our proposed algorithm. We state the theoretical justifications for our algorithm in \textit{Section \ref{Sec:Theoretical_results}}.
The numerical results validating our theory are given in
\textit{Section \ref{Sec:Simulations}}.
We show the application of our model to real data (\cite{winding2023connectome}, \cite{eschbach2020recurrent})
in \textit{Section \ref{Sec:Real_Data}}.
\textit{Section \ref{Sec:Discussion}} concludes by discussing certain recommendations in specific cases of deviation from our model assumptions and some possible future extensions. 
Finally, the proofs of our theoretical results are given in the supplemental materials.
\section{Important definitions and results}
\label{Sec:Definitions_and_notations}
Discussed in this section are some important definitions and notions that we will frequently encounter in this paper.
\subsection{Preliminiaries on
stochastic blockmodels (SBM) and
common subspace independent edge (COSIE)
random graphs}
\label{Subsec:Prelims_SBM_COSIE}
A graph is an ordered pair $(V,E)$ where $V$ is the set of vertices and $E \subset V \times V$ is the set of edges. An adjacency matrix $\mathbf{A}$ of a graph is defined as
$\mathbf{A}_{ij}=1$ if $(i,j) \in E$, and $\mathbf{A}_{ij}=0$ otherwise. Here, we deal with hollow and undirected graphs, hence $\mathbf{A}_{ii}=0$ for all $i$ and is symmetric. Latent position random graphs are those each of whose nodes is associated with a vector that is called its latent position. We denote by $\mathbf{x}_i$ the latent position of the $i$-th node. 
\newline
\newline
First, we state the definition of the common subspace independent edge (COSIE) graph model, from which the graphs in our paper will be sampled.
\begin{definition}
\label{def:COSIE_ordinary}
(\cite{JMLR:v22:19-558})
    Suppose we observe the graphs $G_1,\dots G_N$, or equivalently, their adjacency matrices $\mathbf{A}^{(1)}, \dots \mathbf{A}^{(N)} \in \mathbb{R}^{n \times n}$. 
    We say $(\mathbf{A}^{(1)},\dots \mathbf{A}^{(N)}) \sim \mathrm{COSIE}(\mathbf{V};\mathbf{R}^{(1)},\dots \mathbf{R}^{(N)})$, where $\mathbf{V} \in \mathbb{R}^{n \times d}$ is a matrix of orthonormal columns, and $\mathbf{R}^{(k)} \in \mathbb{R}^{d \times d}$, $k \in [N]$ are symmetric matrices, also known as score matrices, if 
    for all $k \in [N]$,
    $\mathbf{P}^{(k)}= \mathbb{E}(\mathbf{A}^{(k)})=
    \mathbf{V} \mathbf{R}^{(k)} \mathbf{V}
    $, $\mathbf{A}^{(k)}_{ij} \sim \mathrm{Bernoulli}(\mathbf{P}^{(k)}_{ij})$ for $i<j$ and $\mathbf{A}^{(k)}_{ji}=\mathbf{A}^{(k)}_{ij}$.
\end{definition}
In real life, many networks are sparse, meaning the number of edges grow slowly with the number of nodes. To account for sparsity, the definition of COSIE model is modified and stated below. 
\begin{definition}
\label{def:COSIE_sparse}
In the setting of \textit{Definition \ref{def:COSIE_ordinary}},
    suppose we observe adjacency matrices $\mathbf{A}^{(1)},\dots \mathbf{A}^{(N)}$. We say 
    $(\mathbf{A}^{(1)}, \dots \mathbf{A}^{(N)}) \sim \mathrm{COSIE}(\mathbf{V}; \mathbf{R}^{(1)}, \dots \mathbf{R}^{(N)};\rho_n)$ if for all $k \in [N]$,
$\mathbf{P}^{(k)}=\mathbb{E}(\mathbf{A}^{(k)})=
    \rho_n \mathbf{V} \mathbf{R}^{(k)} \mathbf{V}^T$, 
    $\mathbf{A}^{(k)}_{ij} \sim \mathrm{Bernoulli}(\mathbf{P}^{(k)}_{ij})$ 
    for $i<j$ independently and $\mathbf{A}^{(k)}_{ji}=\mathbf{A}^{(k)}_{ij}$, where $\rho_n \to 0$ as $n \to \infty$.
\end{definition}
\begin{remark}
    Observe that setting $\rho_n=1$ in \textit{Definition \ref{def:COSIE_sparse}} recovers \textit{Definition \ref{def:COSIE_ordinary}}.
\end{remark}
\begin{sloppypar}
\begin{remark}
    In our paper, henceforth, whenever we encounter 
    $(\mathbf{A}^{(1)},\dots \mathbf{A}^{(N)}) \sim \mathrm{COSIE}(\mathbf{V};\mathbf{R}^{(1)},\dots \mathbf{R}^{(N)};\rho_n)$,
    we shall assume for all
    $k \in [N]$, for all $i \neq j$, $\left(\mathbf{V} \mathbf{R}^{(k)} \mathbf{V}^T
    \right)_{ij}=\mathbf{V}^T_{i*} \mathbf{R}^{(k)} \mathbf{V}_{j*}=1$. Observe that this can be assumed without loss of generality.
\end{remark}
\end{sloppypar}
We state below the algorithm to estimate the parameters of a sparse common subspace independent edge model.
\begin{algorithm}[H]
\caption{SparseMASE($
(\mathbf{A}^{(1)},\dots \mathbf{A}^{(N)}),d
$)}. 
\label{Algo:MASE_sparse}
\begin{algorithmic}[1]
\State Estimate the sparsity parameter by $\hat{\rho}_n=
\frac{1}{N {n \choose 2}}
\sum_{k=1}^N \sum_{i<j} \mathbf{A}^{(k)}_{ij}
$.
\State For all $k \in [N]$, compute $\hat{\mathbf{V}}^{(k)} \in \mathbb{R}^{n \times d}$, whose columns are the top $d$ left singular vectors of $\mathbf{A}^{(k)}$.
\State Construct 
$\hat{\mathbf{V}}^*=[\hat{\mathbf{V}}^{(1)}| \dots \hat{\mathbf{V}}^{(N)}]$, and compute $\hat{\mathbf{V}} \in \mathbb{R}^{n \times d}$, the matrix whose columns are top $d$ left singular vectors of $\hat{\mathbf{V}}^*$.
\State For all $k \in [N]$, compute $\hat{\mathbf{R}}^{(k)}=
\frac{1}{\hat{\rho}_n}
(\hat{\mathbf{V}}^{(k)})^T \mathbf{A}^{(k)} \hat{\mathbf{V}}^{(k)}$.
\State \Return $(\hat{\mathbf{R}}^{(1)},\dots \hat{\mathbf{R}}^{(N)})$.
\end{algorithmic}
\end{algorithm}
\begin{remark}
    The rationale behind estimating the sparsity parameter in the way mentioned above is justified by results from \cite{agterberg2020nonparametric} and \cite{sekhon2021result}.
\end{remark}
Next, we state the definition of stochastic blockmodels, which comprise a specific category of the common subspace independent edge model. The idea behind stochastic blockmodel is to model networks for which interaction probabilities between nodes depend only upon the communities to which the nodes belong. 
\begin{definition} 
(\cite{Holland1983StochasticBF})
    Suppose the adjacency matrix $\mathbf{A} \in \left\lbrace 0,1 \right\rbrace^{n \times n}$ of an undirected graph with $n$ nodes, satisfies 
    \begin{equation*}
        \mathbb{E}(\mathbf{A})=
        \mathbf{Z} \mathbf{B} \mathbf{Z}^T
    \end{equation*}
where $\mathbf{B} \in [0,1]^{K \times K}$ is symmetric and
$\mathbf{Z} \in \left\lbrace 0,1 \right\rbrace^{n \times K}$
is such that for all $i \in [n]$, $\sum_{j=1}^K \mathbf{Z}_{ij}=1$. Then it is said that the graph is a stochastic blockmodel with  community membership matrix $\mathbf{Z}$ and block connection probability matrix $\mathbf{B}$, and is given by 
$\mathbf{A} \sim \mathrm{SBM}(\mathbf{Z};\mathbf{B})$.
The matrix $\mathbf{Z}$ is such that $\mathbf{Z}_{ik}=1$ if the $i$-th node belongs to the $k$-th community, and $\mathbf{Z}_{ik}=0$ otherwise, $i \in [n]$, $k \in [K]$.  
 The matrix $\mathbf{B}$ is such that $\mathbf{B}_{hk}$ is the probability of formation of an edge between two nodes, one of which belongs to the $h$-th community and the other to the $k$-th community.
\end{definition}
 Secondly, we state the formal definition of multilayer stochastic blockmodel.
\begin{definition}
(\cite{Holland1983StochasticBF})
    Suppose we have $N$ graphs with adjacency matrices $(\mathbf{A}^{(1)}, \dots \mathbf{A}^{(N)}) \in \left\lbrace 0,1 \right\rbrace^{n \times n}$, such that for all $k \in [N]$,
    \begin{equation*}
        \mathbb{E}(\mathbf{A}^{(k)})=
        \mathbf{Z} \mathbf{B}^{(k)} 
        \mathbf{Z}^T
    \end{equation*}
where for all $k \in [N]$,
$\mathbf{B}^{(k)} \in [0,1]^{K \times K}$ is symmetric and 
$\mathbf{Z} \in \left\lbrace 0,1 \right\rbrace^{n \times K}$ is such that for all $i \in [n]$, $\sum_{j=1}^K \mathbf{Z}_{ij}=1$. Then we say that the graphs are jointly distributed as multilayer stochastic blockmodel, represented as 
$(\mathbf{A}^{(1)},\dots \mathbf{A}^{(N)}) \sim \mathrm{MSBM}(\mathbf{Z};\mathbf{B}^{(1)},\dots \mathbf{B}^{(N)})$. Thus,  a multilayer stochastic blockmodel defines a collection of graphs on a common set of vertices, where  every vertex belongs to a unique community irrespective of the layer, but the probability of formation of edge between vertices from different communities changes across layers.
\end{definition}
\begin{remark}
    Common subspace independent edge graph model is a generalization of multilayer stochastic blockmodel, that is, any set of graphs jointly distributed as multilayer stochastic blockmodel can be represented as common subspace independent edge model. If adjacency matrices $(\mathbf{A}^{(1)}, \dots \mathbf{A}^{(N)}) \sim \mathrm{MSBM}(\mathbf{Z};\mathbf{B}^{(1)},\dots \mathbf{B}^{(N)})$
    then $(\mathbf{A}^{(1)},\dots \mathbf{A}^{(N)}) \sim \mathrm{COSIE}(\mathbf{V};\mathbf{R}^{(1)},\dots \mathbf{R}^{(N)})$
    where $\mathbf{V}=\mathbf{Z}(\mathbf{Z}^T \mathbf{Z})^{-\frac{1}{2}}$ and
    for all $k \in [N]$,
    $\mathbf{R}^{(k)}=
    (\mathbf{Z}^T \mathbf{Z})^{\frac{1}{2}}
    \mathbf{B}^{(k)}
    (\mathbf{Z}^T \mathbf{Z})^{\frac{1}{2}}
    $ (for detailed proof, see
    \textit{Appendix A.1 in \cite{JMLR:v22:19-558}}).
\end{remark}
Stochastic blockmodel has an intuitive appeal, where every node belongs to a community and the chance of interaction between two nodes depends on the community memberships of the corresponding nodes. 
Network data arising from different fields of real life can be modeled by multilayer stochastic blockmodel. The common subspace independent edge model is a generalization to multilayer stochastic blockmodel. The common subspace independent edge model is capable of capturing the heterogeneity of real-world multiple network data, while being simple enough to be amenable to algebraic treatments. This is what motivates us to use this model for our study. 
\subsection{Manifold learning by isomap} 
\label{Subsec:Isomap}
Our model involves a sequence of COSIE random graphs, each associated with a scalar response, and each graph corresponding to a point on an unknown one-dimensional manifold in a higher dimensional ambient space. In order to predict the response corresponding to an out-of-sample graph from the same model, we wish to learn the manifold using the procedure isomap (\cite{article}). The problem of manifold learning involves estimating the geodesic distance between a given pair of points on the manifold. Given points $\mathbf{q}_1, \dots \mathbf{q}_n \in \mathcal{M} \subset \mathbb{R}^D$ where $\mathcal{M}$ is a one-dimensional compact Riemannian manifold, the goal is to find scalars $\hat{z}_1,\dots \hat{z}_n$ such that the pairwise interpoint distances between the $\hat{z}_i$ approximate the corresponding pairwise geodesic distances between $\mathbf{q}_i$. The following theorem (\cite{trosset2021rehabilitating}) demonstrates how to estimate the interpoint geodesic distance between a given pair of points on the manifold.
\begin{theorem}
\label{Th:geodesic_by_shortest_path}
(\cite{Bernstein00graphapproximations},\cite{trosset2021rehabilitating})
    Let datapoints $\mathbf{q}_1,\dots \mathbf{q}_n \in \mathbb{R}^D$ be given on  a one-dimensional compact Riemannian manifold $\mathcal{M}$ in ambient space $\mathbb{R}^D$, for which $r_0$ and $s_0$ be the minimum radius of curvature and minimum branch separation respectively. Assume $\nu$ is given and $\lambda>0$ is chosen such that 
    $\lambda<s_0$ and $\lambda<\frac{2}{\pi}r_0 \sqrt{24 \nu}$. Additionally, suppose there 
    exists $\delta>0$ such that
     for every $\mathbf{u} \in \mathcal{M}$, there exists $i \in [n]$ for which $d_M(\mathbf{u},\mathbf{q}_i)<\delta$. A localization graph is constructed on the datapoints $\mathbf{q}_i$ as nodes by the following rule: two points $\mathbf{q}_i$ and $\mathbf{q}_j$ are joined by an edge if $\left\lVert  
     \mathbf{q}_i-\mathbf{q}_j
     \right\rVert<\lambda$. Assuming $\delta<\frac{\nu \lambda}{4}$, the following condition holds for all $i \in [n]$,
     \begin{equation*}
         (1-\nu)d_M(\mathbf{q}_i,\mathbf{q}_j)<d_{n,\lambda}(\mathbf{q}_i,\mathbf{q}_j)<(1+\nu)d_M(\mathbf{q}_i,\mathbf{q}_j)
     \end{equation*}
     where $d_{n,\lambda}(\mathbf{q}_i,\mathbf{q}_j)$ is the shortest path distance between the points $\mathbf{q}_i$ and $\mathbf{q}_j$.
\end{theorem}
Given the dissimilarity matrix $\boldsymbol{\Delta}=
\left(
d_{n,\lambda}(\mathbf{q}_i,\mathbf{q}_j)
\right)_{i,j=1}^n
$, the raw stress at the point 
$(z_1,\dots z_n)$ is defined as
\begin{equation*}
    \sigma(z_1,\dots z_n)=
     \sum_{i,j=1}^n w_{ij}
    \left(
    |z_i-z_j|-d_{n,\lambda}(\mathbf{q}_i,\mathbf{q}_j)
    \right)^2
\end{equation*}
where $w_{ij}$ are weights. 
Setting $w_{ij}=1$, the isomap embeddings are given by
\begin{equation*}
    (\hat{z}_1,\dots \hat{z}_n)=
    \arg \min \sum_{i,j=1}^n
    \left(
    |z_i-z_j|-d_{n,\lambda}(\mathbf{q}_i,\mathbf{q}_j)
    \right)^2.
\end{equation*}
The procedure for isomap (by raw stress minimization) is formally stated in the following algorithm. 
\begin{algorithm}[H]
\caption{ ISOMAP($\left\lbrace \mathbf{q}^{(k)}
\right\rbrace_{k=1}^N,
\lambda,l)$
}
\label{Algo:isomap}
\begin{algorithmic}[1]
\State Construct a localization graph on the points $\mathbf{q}^{(k)}$ as nodes by the following rule: 
join two nodes $\mathbf{q}^{(h)}$ and $\mathbf{q}^{(k)}$ with an edge if $\left\lVert 
\mathbf{q}^{(h)}-\mathbf{q}^{(k)}
\right\rVert<\lambda$.
\State For every $h,k \in [N]$,
compute the shortest path distance $d_{N,\lambda}(\mathbf{q}^{(h)},\mathbf{q}^{(k)})$.
\State Obtain
$(\hat{z}_1,....\hat{z}_l)
= \text{arg min} \sum_{h=1}^l \sum_{k=1}^l (|z_{h}-z_{k}|-d_{N,\lambda}(\mathbf{q}^{(h)},\mathbf{q}^{(k)}))^2$.
\State \Return $(\hat{z}_1,\dots \hat{z}_l)$.
\end{algorithmic}
\end{algorithm}
\begin{remark}
As pointed out in \cite{trosset2021rehabilitating}, isomap operates in two steps: approximating the geodesic distances with shortest path distances and finding low-dimensional embeddings whose pairwise Euclidean distances can well approximate the shortest path distances. Originally in \cite{article}, the second step, that is, the process of finding low-dimensional embeddings for a given dissimilarity matrix of shortest path distances, was proposed to be performed by classical multidimensional scaling. However, \cite{trosset2021rehabilitating} points out that such need not be the only way as there can be other ways to approximate the given dissimilarity matrix of shortest path distances, such as raw stress minimization.    
\end{remark}
\begin{remark}
    The process of minimizing the raw stress function is done by iterative majorization (for details, see Chapter $8$ of \cite{borg2005modern}). Sometimes the algorithm can get trapped in nonglobal minimum, and it can be usually avoided by initializing the algorithm by classical multidimensional scaling outputs. In our paper for theoretical purposes, we assume that the global minima is reached.
\end{remark}
More generally, isomap finds a set of vectors whose pairwise Euclidean distances optimize some loss function with respect to a given dissimilarity matrix. We will now formally generalize this notion. First, we will state a few important definitions. 
\begin{definition}
\label{Def:EDM1}
    A matrix $\mathbf{D} \in \mathbb{R}^{l \times l}$ is called EDM-1 if there exists $p \in \mathbb{N}$ and points $\mathbf{z}_1,\dots \mathbf{z}_l \in \mathbb{R}^p$ such that for all $i,j \in [l]$, $\mathbf{D}_{ij}=\left\lVert \mathbf{z}_i-\mathbf{z}_j \right\rVert$.
    The smallest such $p$ is called the embedding dimension of $\mathbf{D}$.
\end{definition}
Then, given the dissimilarity matrix $\boldsymbol{\Delta}$ of shortest path distances, isomap solves the problem of $\arg \min_{\mathbf{D} \in \mathscr{Y}_l} \left\lVert \mathbf{D}-\boldsymbol{\Delta} \right\rVert$ where $\mathscr{Y}_l$ denotes the closed cone of all $l \times l$ EDM-1 matrices of embedding dimension less than or equal to $d$.
\newline
\newline
The next section describes in detail our model and the proposed algorithm.
\section{Model and Methodology} 
\label{Sec:Model_and_Method}
Our model involves a sequence of common subspace independent edge random graphs $(\mathbf{A}^{(1)},\dots \mathbf{A}^{(N)}) \sim \mathrm{COSIE}(\mathbf{V};\mathbf{R}^{(1)},\dots \mathbf{R}^{(N)};\rho_n)$ where
$\mathbf{V} \in \mathbb{R}^{n \times d}$ is the common subspace matrix and  $\mathbf{R}^{(k)} \in \mathbb{R}^{d \times d}$ are the symmetric score matrices. 
In our model, we assume $d$ to be known. 
By definition, the columns of $\mathbf{V}$ are orthonormal vectors. The probability matrices are given by $\mathbf{P}^{(k)}=\rho_n \mathbf{V} \mathbf{R}^{(k)} \mathbf{V}^T$,
$k \in [N]$, where $\rho_n$ is the sparsity parameter satisfying $\rho_n \to 0$ as $n \to \infty$. The following assumptions are made about our model.
\begin{assumption}
\label{Asm:Delocalization}
    There exist constants $c_1,c_2$ and an orthogonal matrix $\mathbf{W} \in \mathcal{O}(d)$, such that for every $i \in [n], j \in [d]$, 
    \begin{equation*}
        \frac{c_1}{\sqrt{n}}
        < (\mathbf{V} \mathbf{W})_{ij}< 
        \frac{c_2}{\sqrt{n}}.
    \end{equation*}
\end{assumption}
The above assumption ensures 
that the score matrices influence the connectivity of enough edges in the graph, and
(\textit{Assumption \ref{Asm:Delocalization}}) is satisfied by various networks, for instance by Erdos-Renyi graphs and by stochastic blockmodels whose community sizes grow linearly with the number of nodes (\cite{JMLR:v22:19-558}).
\begin{assumption}
\label{Asm:Edge_variance}
    For all $k \in [N]$,
    \begin{equation*}
        s^2(\mathbf{P}^{(k)})=
        \sum_{i,j=1}^n \mathbf{P}^{(k)}_{ij} 
        (1-\mathbf{P}^{(k)}_{ij})=
        \omega(1) 
    \end{equation*}
    and for $p \neq q$,
    \begin{equation*}
        \sum_{i,j=1}^n \mathbf{P}^{(k)}_{ij} 
        (1-\mathbf{P}^{(k)}_{ij})
        (
        n \mathbf{V}_{ip} \mathbf{V}_{jq} +
        n \mathbf{V}_{jp} \mathbf{V}_{iq}
        )^2=\omega(1).
    \end{equation*}
\end{assumption}
A balanced multilayer stochastic blockmodel for which the edge formation probabilities grow at par with the number of nodes, will satisfy the above condition (\textit{Assumption \ref{Asm:Edge_variance}}).
\begin{assumption}
\label{Asm:score_matrix_variance}
    Define
    \begin{equation*}
\boldsymbol{\Sigma}^{(k)}_{
\frac
{2p+q(q-1)}{2},
\frac{2r+s(s-1)}{2}
}=
        \sum_{i=1}^{n-1} \sum_{j=i+1}^n
        \mathbf{P}^{(k)}_{ij}
        (1-\mathbf{P}^{(k)}_{ij})
        (\mathbf{V}_{ip} \mathbf{V}_{jq} +
        \mathbf{V}_{jp} 
        \mathbf{V}_{iq})
        (\mathbf{V}_{ir} \mathbf{V}_{js} +
        \mathbf{V}_{jr} 
        \mathbf{V}_{is}).
    \end{equation*}
    Then, for all $k \in [N]$,
    \begin{equation*}
        \lambda_{min}(\boldsymbol{\Sigma}^{(k)})=\omega(n^{-2}).
    \end{equation*}
\end{assumption}
The assumption stated above (\textit{Assumption \ref{Asm:score_matrix_variance}}) serves as a sufficient condition to ensure that the joint distribution of the entries in upper triangle of the estimated score matrices can be derived, which in turn leads to the key result of \cite{JMLR:v22:19-558}, upon which our results are pivoted. 
\begin{assumption}
    \label{Asm:largest_expected_degree_growth}
    We assume
    \begin{equation*}
        \epsilon=
        \sqrt{
        \frac{1}{\rho_n^2 N}
        \sum_{k=1}^N
        \frac
        {
        \delta(\mathbf{P}^{(k)})
        }
        {
        \lambda_{min}^2 (\mathbf{R}^{(k)})
        }
        }
        =o(1)
    \end{equation*}
    and 
    \begin{equation*}
    \min_{k \in [N]}
        \delta(\mathbf{P}^{(k)})=
        \omega(\mathrm{log}(n)).
    \end{equation*}
\end{assumption}
The condition stated above controls the sparsity of the graphs, and networks that are extremely sparse will fail to satisfy \textit{Assumption \ref{Asm:largest_expected_degree_growth}}.
\begin{assumption}
\label{Asm:scaled_R_free_of_graph_size}
    It is assumed that for all $k$, the quantity $\frac{1}{n} \mathbf{R}^{(k)}$ does not depend on $n$.
\end{assumption}
Balanced multilayer stochastic blockmodels will satisfy \textit{Assumption \ref{Asm:scaled_R_free_of_graph_size}}, and a multilayer stochastic blockmodel where the growth of community sizes vary widely from one another will fail to satisfy it.  
\newline
\newline
Define the scaled score matrices to be $\mathbf{Q}^{(k)}=\frac{1}{n} \mathbf{R}^{(k)}$ and subsequently
define $\mathbf{q}^{(k)}=\mathrm{vec}(\mathbf{Q}^{(k)}) \in \mathbb{R}^D$
for $k \in [N]$, 
where $D=d^2$. For all $k$, it is assumed that the vectors $\mathbf{q}^{(k)} \in \mathcal{M}$, where $\mathcal{M}=\psi([0,L])$ is a one-dimensional compact Riemannian manifold in ambient space $\mathbb{R}^D$, $\psi:[0,L] \to \mathbb{R}^D$ being a bijective and sufficiently well-behaved function. Denote the scalar pre-image of $\mathbf{q}^{(k)}$ by $t_k$ for all $k \in [N]$, that is,
$\mathbf{q}^{(k)}=\psi(t_k)$.
Suppose for all
$k \in [s]$, where $s \ll N$ is fixed,
$\mathbf{A}^{(k)}$ is associated with response $y_k$, and the following regression model is assumed to hold:
\begin{equation*}
    y_k=\alpha+\beta t_k +\epsilon_k
\end{equation*}
where $\epsilon_k \sim^{iid} N(0,\sigma^2_{\epsilon})$ for $k \in [s]$. Our goal is to predict the response $y_r$ corresponding to the graph $\mathbf{A}^{(r)}$, $r>s$. 
The procedure is stated in the following algorithm.
\begin{algorithm}[H]
\caption{PredGraphResp($
\left\lbrace\mathbf{A}^{(k)}
\right\rbrace_{k=1}^N,
 \left\lbrace y_k \right\rbrace_{k=1}^s,
d,\lambda,l,N^*,
r
$)}
\label{Algo:predict_response}
\begin{algorithmic}[1]
\State Obtain the estimates (upto orthogonal rotation)
of the score matrices,
$(\hat{\mathbf{R}}^{(1)},\dots \hat{\mathbf{R}}^{(N)})=\mathrm{SparseMASE}(\mathbf{A}^{(1)},\dots \mathbf{A}^{(N)},d)$, and compute
$\hat{\mathbf{Q}}^{(k)}=\frac{1}{n}\hat{\mathbf{R}}^{(k)}$, and subsequently
$\hat{\mathbf{q}}^{(k)}=\mathrm{vec}(\hat{\mathbf{Q}}^{(k)})$ for all $k \in [N]$.
\State Construct a localization graph with $\left\lbrace
\hat{\mathbf{q}}^{(k)}
\right\rbrace_{k=1}^{N^*}$ as vertices by the following rule: join two vertices $\hat{\mathbf{q}}^{(h)},\hat{\mathbf{q}}^{(k)}$ if
and only if $\left\lVert \hat{\mathbf{q}}^{(h)}-\hat{\mathbf{q}}^{(k)}
\right\rVert<\lambda$. 
\State For every $h,k \in [N^*]$, compute the shortest path distance $d_{N^*,\lambda}(\hat{\mathbf{q}}^{(h)},\hat{\mathbf{q}}^{(k)})$. 
\State Obtain
$(\hat{z}_1,....\hat{z}_l)
= \text{arg min} \sum_{h=1}^l \sum_{k=1}^l (|z_{h}-z_{k}|-d_{N^*,\lambda}(\hat{\mathbf{q}}^{(h)},\hat{\mathbf{q}}^{(k)}))^2$.
\State Compute 
$\hat{b}=\frac
{
\sum_{i=1}^s (y_i-\Bar{y})(\hat{z}_i-\Bar{\hat{z}})
}
{
\sum_{i=1}^s (\hat{z}_i-\Bar{\hat{z}})^2
}$ and
$\hat{a}=\Bar{y}-\hat{b} \Bar{\hat{z}}$.
\State  Compute $\Tilde{y}_r=\hat{a}+\hat{b} \hat{z}_r$.
\State \Return $\Tilde{y}_r$. 
\end{algorithmic}
\end{algorithm}
\begin{remark}
    If the graphs are sampled from a balanced multilayer stochastic blockmodel (that is, a multilayer stochastic blockmodel where a node is equally likely to belong to any of the communities), then the scaled score matrices are same as the block connection probability matrices. In this particular setting our model basically assumes that the block connection probability matrices are functions of some scalar values, and the responses are linked to these scalar pre-images via a simple linear regression model.
\end{remark}
Next, we discuss the theoretical results that justify the use of our proposed \textit{Algorithm \ref{Algo:predict_response}}.
\section{Theoretical results}
\label{Sec:Theoretical_results}
In this section, we state our key theoretical results. Our results are pivoted primarily on two results in the literature: asymptotic normality of the estimated score matrices by multiple adjacency spectral embedding (\cite{JMLR:v22:19-558}) and 
a continuity theorem on raw-stress embeddings (\cite{trosset2024continuous}).
From the following result in \cite{JMLR:v22:19-558}, we know that the estimates $\hat{\mathbf{R}}^{(k)}$ exhibit an asymptotic normality property. 
\begin{theorem}
(\cite{JMLR:v22:19-558})
\label{Th:MASE_R_asy}
    Suppose adjacency matrices 
    $(\mathbf{A}^{(1)}, \dots \mathbf{A}^{(N)}) \sim \mathrm{COSIE}(\mathbf{V}; \mathbf{R}^{(1)}, \dots \mathbf{R}^{(N)};\rho_n)$, where 
    $\mathbf{V} \in \mathbb{R}^{n \times d}$, $\mathbf{R}^{(k)} \in \mathbb{R}^{d \times d}$ for all $k \in [N]$ and the sparsity parameter $\rho_n \to 0$ as $n \to \infty$.
     Assume that \textit{Assumptions \ref{Asm:Delocalization}, \ref{Asm:Edge_variance},
    \ref{Asm:score_matrix_variance},
\ref{Asm:largest_expected_degree_growth},
\ref{Asm:scaled_R_free_of_graph_size}
} hold. 
    Then there exists a sequence of matrices $\mathbf{W} \in \mathcal{O}(d)$, such that for all $k \in [N]$ and
    $i,j \in [d]$, as $n \to \infty$,
    \begin{equation*}
        \frac{1}{\sigma_{k,i,j}}\left(
      \hat{\rho}_n  \hat{\mathbf{R}}^{(k)}-
      \rho_n
        \mathbf{W}^T \mathbf{R}^{(k)} 
        \mathbf{W} +
        \mathbf{H}^{(k)}
        \right)_{ij}
        \to^d N(0,1),
    \end{equation*}
    where $\mathbb{E} (
    \left\lVert \mathbf{H}^{(k)}
    \right\rVert_F)=
    O(\frac{d}{\sqrt{N}})
    $ and $
    \sigma^2_{k,i,j}=O(1)
    $.
\end{theorem}
From the above theorem \ref{Th:MASE_R_asy}, it can be established that $\hat{\mathbf{Q}}^{(k)}=\frac{1}{n} \hat{\mathbf{R}}^{(k)}$ estimates $\mathbf{Q}^{(k)}=\frac{1}{n} \mathbf{R}^{(k)}$ consistently as $n \to \infty, N \to \infty$ under appropriate regularity assumptions, upto an orthogonal transformation. This enables us to establish that the pairwise Frobenius distances among a fixed set of $\hat{\mathbf{Q}}^{(k)}$ will consistently estimate the corresponding Frobenius distances among the $\mathbf{Q}^{(k)}$ (see \textit{Lemma $2$}). This, in turn, leads us to conclude that pairwise shortest path distances in a localization graph constructed on $\hat{\mathbf{q}}^{(k)}$ as vertices approach the corresponding pairwise shortest path distances in a localization graph constructed on $\mathbf{q}^{(k)}$ as vertices, as total number of graphs go to infinity. This argument is formally stated below.
\begin{sloppypar}
\begin{proposition}
\label{Prop:shortest_path_distance_convergence}
Let $N^* \in \mathbb{N}$ be fixed and
    suppose $(\mathbf{A}^{(1)},\dots \mathbf{A}^{(N^*)},\dots \mathbf{A}^{(N)}) \sim \mathrm{COSIE}(\mathbf{V};\mathbf{R}^{(1)},\dots \mathbf{R}^{(N^*)},\dots \mathbf{R}^{(N)};\rho_n)$ where
    $\mathbf{V} \in \mathbb{R}^{n \times d}$ is the common subspace matrix, $\rho_n \mathbf{R}^{(k)}, k \in [N]$ are the symmetric score matrices where $\rho_n \to 0$ as $n \to \infty$, and let $\mathbf{q}^{(k)}=\mathrm{vec}(\frac{1}{n}\mathbf{R}^{(k)})$ for all $k \in [N]$. Denote the multiple adjacency spectral embedding outputs by $(\hat{\mathbf{R}}^{(1)}, \dots \hat{\mathbf{R}}^{(N)})=\mathrm{SparseMASE}(\mathbf{A}^{(1)},\dots \mathbf{A}^{(N)},d)$ and subsequently define $\hat{\mathbf{q}}^{(k)}=\mathrm{vec}(\frac{1}{n}\hat{\mathbf{R}}^{(k)})$ for $k \in [N]$. Assume that $d_{\hat{G},\lambda}(\hat{\mathbf{q}}^{(h)},\hat{\mathbf{q}}^{(k)})$ denotes the shortest path distance between $\hat{\mathbf{q}}^{(h)}$ and $\hat{\mathbf{q}}^{(k)}$ in the localization graph $\hat{G}$ with neighbourhood parameter $\lambda$ constructed on the points $\left\lbrace \hat{\mathbf{q}}^{(k)}\right\rbrace_{k=1}^{N^*}$, and
    $d_{\Tilde{G},\lambda}(\mathbf{q}^{(h)},\mathbf{q}^{(k)})$ denotes the shortest path distance between $\mathbf{q}^{(h)}$ and $\mathbf{q}^{(k)}$ in the localization graph $\Tilde{G}$ with
    neighbourhood parameter $\lambda$
    constructed on the points $\left\lbrace \mathbf{q}^{(k)}\right\rbrace_{k=1}^{N^*}$. Define
    \begin{equation*}
    \hat{\boldsymbol{\Delta}}=
    \left(
    d_{\hat{G},\lambda}
    (\hat{\mathbf{q}}^{(h)},
    \hat{\mathbf{q}}^{(k)})
    \right)_{h,k=1}^l,
    \hspace{0.2cm}
    \Tilde{\boldsymbol{\Delta}}=
    \left(
    d_{\Tilde{G},\lambda}
    (\mathbf{q}^{(h)},\mathbf{q}^{(k)})
    \right)_{h,k=1}^l \in \mathbb{R}^{l \times l}.
\end{equation*}
For any $\lambda>0$, as $n \to \infty$ and $N \to \infty$, 
\begin{equation*}
    \left\lVert 
    \hat{\boldsymbol{\Delta}}-
    \Tilde{\boldsymbol{\Delta}}
    \right\rVert_F \to 0.
\end{equation*}
\end{proposition}
\end{sloppypar}
Recall that from \textit{Theorem \ref{Th:geodesic_by_shortest_path}}, we know that the pairwise shortest path distances between the points $\mathbf{q}^{(k)}$ converge to the corresponding geodesic distances on a sequence of appropriately constructed localization graphs. Hence, \textit{Proposition \ref{Prop:shortest_path_distance_convergence}} leads us
to infer that the pairwise shortest path distances between the points $\hat{\mathbf{q}}^{(k)}$
approach the corresponding true geodesic distances between the points $\mathbf{q}^{(k)}$, which is formally stated below.
\begin{proposition}
\label{Prop:MASE_output_shortest_path_to_true_geodesics}
Let $l \in \mathbb{N}$ be fixed.
Using the notation from \textit{Proposition \ref{Prop:shortest_path_distance_convergence}},
    suppose 
    \newline
    $(\mathbf{A}^{(1)},\dots \mathbf{A}^{(N^*)},\dots \mathbf{A}^{(N)}) \sim \mathrm{COSIE}(\mathbf{V};\mathbf{R}^{(1)},\dots \mathbf{R}^{(N^*)},\dots \mathbf{R}^{(N)};\rho_n)$. Assume that for all $k$ the points $\mathbf{q}^{(k)}$ lies on an one-dimensional non-self-intersecting compact Riemannian manifold $\mathcal{M}$, and let $d_M(\mathbf{q}^{(h)},\mathbf{q}^{(k)})$ denote the geodesic distance between the points $\mathbf{q}^{(h)}$ and $\mathbf{q}^{(k)}$.
    Additionally, assume that \textit{Assumptions
    \ref{Asm:Delocalization},
    \ref{Asm:Edge_variance},
    \ref{Asm:score_matrix_variance},
    \ref{Asm:largest_expected_degree_growth},
\ref{Asm:scaled_R_free_of_graph_size}
    } hold.
    There exist sequences 
    $\left\lbrace \lambda_K \right\rbrace_{K=1}^{\infty}$ of neighbourhood parameters,
    $\left\lbrace n_K \right\rbrace_{K=1}^{\infty}$ of graph size, $\left\lbrace N_K \right\rbrace_{K=1}^{\infty}$ 
    of total number of graphs
    and 
    $\left\lbrace N^*_K\right\rbrace_{K=1}^{\infty}
    $ of number of graphs for isomap,
    satisfying $\lambda_K \to 0, n_K \to \infty,
    N^*_K \to \infty,
    N_K \to \infty,
    N_K=\omega(N^*_K)
    $ as $K \to \infty$,
    such that when $K \to \infty$,
    \begin{equation*}
        \left\lVert 
        \hat{\boldsymbol{\Delta}}-
        \boldsymbol{\Delta}
        \right\rVert \to 0
    \end{equation*}
    where $\boldsymbol{\Delta}=\left(
    d_M(\mathbf{q}^{(h)},\mathbf{q}^{(k)})
    \right)_{h,k=1}^l$ and
    $
    \hat{\boldsymbol{\Delta}}=
    d_{N^*_K,\lambda_K}
    (\hat{\mathbf{q}}^{(h)},
    \hat{\mathbf{q}}^{(k)})
    $.
\end{proposition}
So far, we are able to prove that the dissimilarity matrix
$\hat{\boldsymbol{\Delta}}=
\left(
d_{\hat{G},\lambda}(\hat{\mathbf{q}}^{(h)},\hat{\mathbf{q}}^{(k)})
\right)_{h,k=1}^l
$
of shortest path distances obtained from the vectorized scaled estimated score matrices $\hat{\mathbf{q}}^{(k)}$ approach the dissimilarity matrix $\boldsymbol{\Delta}
=
\left(
d_M(\mathbf{q}^{(h)},\mathbf{q}^{(k)})
\right)_{h,k=1}^l
$ of the true geodesic distances between the images
$\mathbf{q}^{(k)}=\psi(t_k)$
of the regressors $t_k$. We wish to use this to show that the pairwise distances between the isomap embeddings obtained from the vectorized scaled estimated score matrices 
$\hat{\mathbf{q}}^{(k)}$
approach the pairwise distances between the true regressors $t_k$, which are equal to pairwise geodesic distances between the corresponding images owing to the fact that the manifold $\mathcal{M}=\psi([0,L])$ is arclength-parameterized. We need a continuity theorem to establish our argument, which is provided by  the following theorem (\cite{trosset2024continuous}) which enables us to use the above argument to establish the consistency of isomap embeddings.
\begin{theorem}
\label{Th:dissimilarity_continuity}
\cite{trosset2024continuous})
    Let $\left\lbrace 
    \boldsymbol{\Delta}^{(K)}
    \right\rbrace_{K=1}^{\infty}$ be a sequence of dissimilarity matrices such that for each $K \in \mathbb{N}$, $\boldsymbol{\Delta}^{(K)} \in \mathbb{R}^{l \times l}$, and $\lim_{K \to \infty}
    \left\lVert
    \boldsymbol{\Delta}^{(K)} -\boldsymbol{\Delta}^{(0)}
    \right\rVert_F = 0$. 
     For any dissimilarity matrix $\boldsymbol{\Delta} \in \mathbb{R}^{l \times l}$,
     define the set of globally minimizing EDM-1 matrices for $\boldsymbol{\Delta}$ to be
    $\mathrm{Min}(\boldsymbol{\Delta})=
    \left\lbrace
    \mathbf{D} \in \mathscr{Y}_l:
    \sigma_l(\mathbf{D},\boldsymbol{\Delta})=
    \inf_{\mathbf{D} \in \mathscr{Y}_l}
    \sigma_l(\mathbf{D},\boldsymbol{\Delta})
    \right\rbrace
    $ where $\mathscr{Y}_l$ is the closed cone of all EDM-1 matrices of embedding dimension one.
    If for all $k \in \mathbb{N}$, $\mathbf{D}^{(K)} \in \mathrm{Min}(\boldsymbol{\Delta}^{(K)})$,
    then the sequence $\left\lbrace 
    \mathbf{D}^{(K)}
    \right\rbrace_{K=1}^{\infty}$
has an accumulation point $\mathbf{D}^{(0)}$ such that
$\mathbf{D}^{(0)} \in \mathrm{Min}(\boldsymbol{\Delta}^{(0)})$.
\end{theorem}
We use the abovementioned \textit{Theorem \ref{Th:dissimilarity_continuity}} to prove that the pairwise distances between the isomap embeddings $\hat{z}_k$ obtained from the $\hat{\mathbf{q}}^{(k)}$ approach the corresponding geodesic distances between the $\mathbf{q}^{(k)}$, which equal pairwise distances between the true regressors $t_k$. Finally, using the above arguments, we establish the convergence guarantee for the predicted response obtained from the isomap embeddings, which is formally stated below in \textit{Theorem \ref{Th:isomap_consistency}}.
\begin{sloppypar}
\begin{theorem}
\label{Th:isomap_consistency}
Suppose we have $N$ graphs with adjacency matrices 
$(\mathbf{A}^{(1)}, \dots \mathbf{A}^{(N)}) \sim \mathrm{COSIE}(\mathbf{V};\mathbf{R}^{(1)}, \dots \mathbf{R}^{(N)},\rho_n)$. 
Define $\mathbf{q}^{(k)}=\mathrm{vec}(\mathbf{Q}^{(k)})$ where $\mathbf{Q}^{(k)}=\frac{1}{n} \mathbf{R}^{(k)}$, and assume for all $k$, $\mathbf{q}^{(k)}=\psi(t_k)$ lies on the one-dimensional manifold $\mathcal{M}=\psi([0,L])$. 
Let $s \ll N$ be fixed and responses $y_1, \dots y_s$ are recorded at the first $s$ graphs, and assume the following regression model holds:
\begin{equation*}
    y_i=\alpha+\beta t_i + \epsilon_i 
\end{equation*}
where $\epsilon_i \sim^{iid} N(0,\sigma^2_{\epsilon})$ for $i \in [s]$.
Suppose \textit{Assumptions \ref{Asm:Delocalization}, \ref{Asm:Edge_variance},
\ref{Asm:score_matrix_variance},
\ref{Asm:largest_expected_degree_growth},
\ref{Asm:scaled_R_free_of_graph_size}}
hold. Denote by $\hat{\mathbf{q}}^{(k)}=\mathrm{vec}(\hat{\mathbf{Q}}^{(k)})$, where $\hat{\mathbf{Q}}^{(k)}=\frac{1}{n}\hat{\mathbf{R}}^{(k)}$. There exists a sequence $N_K$ of total number of graphs,
$N^*_K=o(N_K)$ of number of graphs for isomap,
and $\lambda_K$ of neighbourhood parameters, for which $N^*_K \to \infty$, $N_K \to \infty$ and $\lambda_K \to 0$ as $K \to \infty$, such that 
for a fixed $l \in \mathbb{N}$, the predicted response $\Tilde{y}_r=\mathrm{PredGraphResp}(\left\lbrace\mathbf{A}^{(k)}\right\rbrace_{k=1}^{N_K}, 
\left\lbrace y_k \right\rbrace_{k=1}^s,d,
\lambda_K,N^*_K,l,r)$ (see \textit{Algorithm \ref{Algo:predict_response}})
will satisfy: for every $r \leq l$, as $K \to \infty$,
\begin{equation*}
|\Tilde{y}_r-\hat{y}_r| \to^P 0
\end{equation*}
where $\hat{y}_r$ is the predicted response for the $r$-th network based on the true regressors $t_k$.
\end{theorem}
\end{sloppypar}
Thus, in the absence of the true regressors, the isomap embeddings can be used as proxy regressors to predict the responses, the convergence guarantee for which is established in the above \textit{Theorem \ref{Th:isomap_consistency}}.
In order to test the validity of the simple linear regression model $y_k=\alpha+\beta t_k +\epsilon_k,  k \in [s]$ where $\epsilon_k \sim^{iid} N(0,\sigma^2_{\epsilon})$, we conduct statistical hypothesis testing of $H_0: \beta=0$ versus $H_1: \beta \neq 0$. We use a test statistic that depends on the observed responses $y_k$ and the predicted responses $\hat{y}_k$ based on the true regressors $t_k$. However, in the absence of the true regressors $t_k$, we can use their approximations $\Tilde{y}_k$, the predicted responses based on the isomap embeddings $\hat{z}_k$. The corollary stated below,
as a direct consequence of \textit{Theorem \ref{Th:isomap_consistency}},
justifies this argument by formally establishing that the power of the test involving the predicted responses based on the isomap embeddings approach the power of the test based on the true regressors.
\begin{corollary}
\label{Cor:power_convergence}
    In the setting of \textit{Theorem \ref{Th:isomap_consistency}}, suppose we are to conduct the test $H_0:\beta=0$ against $H_1:\beta \neq 0$ at level of significance $\Tilde{\alpha}$. Define the following test statistics:
    \begin{equation*}
        F^*=(s-2)
        \frac
        {
        \sum_{k=1}^s (\hat{y}_k-\Bar{y})^2
        }
        {
        \sum_{k=1}^s (y_k-\hat{y}_k)^2
        }, \hspace{0.5cm}
        \hat{F}=
        (s-2)
        \frac
        {
        \sum_{k=1}^s (\Tilde{y}_k-\Bar{y})^2
        }
        {
        \sum_{k=1}^s
        (y_k-\Tilde{y}_k)^2
        }.
    \end{equation*}
    Suppose $\pi^*$ is the power of the test done by the rule: reject $H_0$ if $F^*>c_{\Tilde{\alpha}}$, and let $\hat{\pi}$ be the power of the test done by the rule: reject $H_0$ if 
    $\hat{F}>c_{\Tilde{\alpha}}$. Then, for every $(\alpha ,\beta)$,
    $|\hat{\pi}-\pi^*| \to 0$ as $K \to \infty$.
\end{corollary}
Thus, we can test the validity of the regression model using the isomap embeddings with power approximately same as that of the test that uses the true regressors. 
\begin{remark}
    Our paper essentially establishes that if the vectors $\mathbf{q}^{(k)}=\mathrm{vec}(\frac{1}{n}\mathbf{R}^{(k)})$ 
    lie on a sufficiently well-behaved one-dimensional manifold, then the isomap embeddings obtained from 
    $\left\lbrace \hat{\mathbf{q}}^{(k)} \right\rbrace$ can be used as proxy regressors. Our key result is pivoted on \textit{Propositions 
\ref{Prop:shortest_path_distance_convergence} and
\ref{Prop:MASE_output_shortest_path_to_true_geodesics}}, which establish that the pairwise shortest path distances of $\left\lbrace \hat{\mathbf{q}}^{(k)}
    \right\rbrace$ approach the pairwise geodesic distances between $\left\lbrace \mathbf{q}^{(k)}
    \right\rbrace$. If, $\mathcal{I} \subset [d] \times [d]$, and if the vectors $\mathbf{q}^{(k)}_{\mathcal{I}}=\mathrm{vec}(\frac{1}{n} \mathbf{R}^{(k)}_{\mathcal{I}})$ satisfy the criterion of lying on a one-dimensional manifold instead of the vectors $\mathbf{q}^{(k)}$, then similar results will hold: the pairwise shortest path distances between the 
    MASE outputs
    $\left\lbrace 
    \hat{\mathbf{q}}^{(k)}_{\mathcal{I}}
    \right\rbrace$ will approach the pairwise geodesic distances between $\left\lbrace 
    \mathbf{q}^{(k)}_{\mathcal{I}}
    \right\rbrace$.
\end{remark}
\begin{remark}
    We draw the attention of the reader to the fact that to apply our algorithm, we are requiring two sets of auxiliary graphs, one set to help us learn the manifold well and another set to help us estimate the points on the manifold with vanishing error. This makes our process somewhat wasteful, although the reason lies in the fact that in \textit{Theorem \ref{Th:MASE_R_asy}}, the bound on the [Frobenius] norm of the bias matrix is pointwise. Had it been a uniform bound, we could have worked with one set of auxiliary graphs. Alternatively, if an improvement over isomap is provided that offers us uniform bound instead of pointwise bound on the error of estimating pairwise geodesic distances, that would also help us work with one, instead of two, set of auxiliary graphs. 
\end{remark}
Having theoretically justified our method, we now move on to the next section to discuss the  numerical results we obtained that support our theory. 
\section{Simulations}
\label{Sec:Simulations}
\begin{figure}[h!]
    \centering
\includegraphics[scale=0.85]{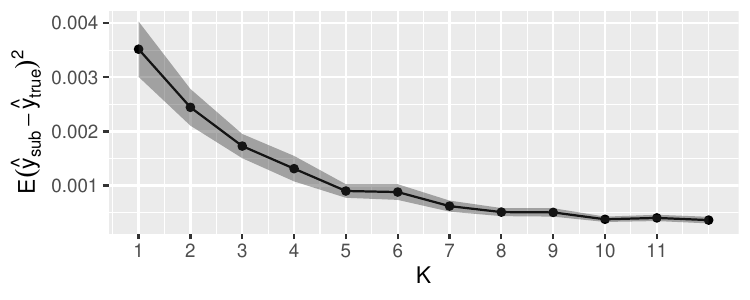}
    \caption{Plot showing the mean squared difference between predicted response based on the isomap embeddings and the predicted response based on the true regressors approach zero as the number of graphs and the number of nodes in each graph increases. For every term in a sequence of number of graphs and size of the graphs, $100$ Monte Carlo samples are generated from a balanced multilayer stochastic blockmodel, which is a special case of COSIE model, and the scaled score matrices are estimated to approximate the points on the manifold. Subsequently isomap is used to learn the manifold and the embddings are used as proxy regressors to predict the response at an unlabeled graph. The plot shows that the mean squared difference between the predicted responses based on the isomap embeddings and the true regressors approach zero.}
    \label{fig:response_consistency}
\end{figure}

\begin{figure}[h!]
    \centering
    \includegraphics[scale=0.85]{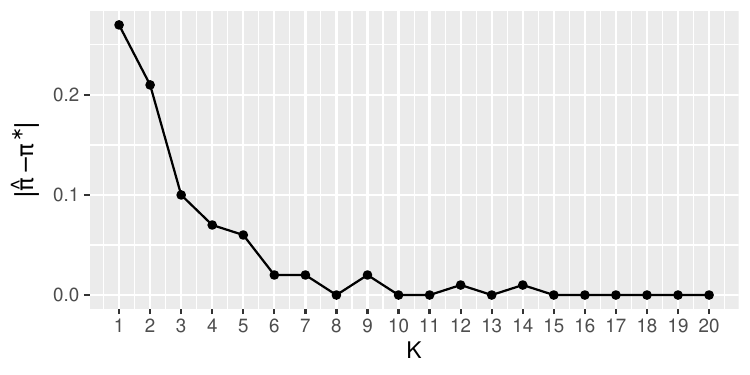}
    \caption{Plot showing the difference between the empirical powers of the tests going to zero as the number of graphs and number of nodes in each graph go to infinity. As before, $100$ Monte Carlo samples are generated from a balanced multilayer stochastic blockmodel, which is a special case of COSIE model, and the scaled score matrices are estimated to approximate the points on the manifold, and isomap is applied to obtain the scalar embeddings which are subsequently used as proxy regressors. The approximate $F$-statistic is computed from the embeddings and finally, the powers of the tests based on the true and the approximate $F$-statistic are empirically estimated and compared.}
    \label{fig:power_convergence}
\end{figure}

In this section, we describe our simulation experiment. We conduct a simulation experiment to provide numerical support for \textit{Theorem \ref{Th:isomap_consistency}}.
We take the number of labeled graphs to be fixed at $s=5$,  set the regression parameters to be $\alpha=2.0$,
$\beta=5.0$ and $\sigma_{\epsilon}=0.01$,
and choose $r=6$.
We define the manifold to be $\mathcal{M}=\psi[0.25,1]$ where $\psi:\mathbb{R} \to \mathbb{R}^D$ is defined as $\psi(t)=(t/a,t/b,t/b,t/a)$, 
where $a=\frac{\sqrt{2}}{\mathrm{sin}(1)}$, $b=\frac{\sqrt{2}}{\mathrm{cos}(1)}$.
We define an index $K$ such that the total number of graphs $N_K \to \infty$,
the number of graphs for isomap $N^*_K \to \infty$,
the size of each graph $n_K \to \infty$ and
the neighbourhood parameter $\lambda_K \to 0$
as $K \to \infty$. We vary $K$ over the range $\left\lbrace 1,2,\dots 12 \right\rbrace$, and set $n_K=500+150(K-1)$, $N_K=15+(K-1)$, 
$N^*_K=
\left\lfloor 
N_K^{\frac{3}{4}}
\right\rfloor
$ and $\lambda_K=2.0 \times 0.99^{K-1}$. For every $K$, we generate $100$ Monte Carlo samples of random graphs and perform the following procedure on each sample. We generate $t_1,\dots t_s,t_{s+1},\dots t_{N_K} \sim^{iid} \mathrm{Uniform}(0.25,1)$ and set $\mathbf{t}=(t_1,\dots t_{N_K})$. For each $t_i$, we form the matrix $\mathbf{B} \equiv \mathbf{B}(t_i) \in \mathbb{R}^{d \times d}$ whose diagonal elements are $t_i/a$ and the off-diagonal elements are $t_i/b$. We also form the common community membership matrix $\mathbf{Z} \in \mathbb{R}^{n_K \times d}$ by the following rule: the first $n_K/2$ rows of $\mathbf{Z}$ are all $(1,0)^T$ and the rest $n_K/2$ rows are all $(0,1)^T$.
Then we form $N_K$ probability matrices $\mathbf{P}(t_i)=\mathbf{Z} \mathbf{B}(t_i) \mathbf{Z}^T$, and we sample $N_K$ symmetric adjacency matrices  $\mathbf{A}^{(k)}$ such that
for all $k \in [N_K]$, and for all $i<j, i,j \in [n]$,
$\mathbf{A}^{(k)}_{ij} \sim \mathrm{Bernoulli}(\mathbf{P}^{(k)}_{ij})$.
. We know that COSIE model is a generalized version of stochastic blockmodel with fixed community memberships, hence $\mathbf{A}^{(1)}, \dots \mathbf{A}^{(N_K)} \sim \mathrm{COSIE}(\mathbf{V};\mathbf{R}^{(1)}, \dots \mathbf{R}^{(N_K)})$ for some $\mathbf{V}$ with orthonormal columns and some $\mathbf{R}^{(k)}$, $k \in [N_K]$. We then compute $(\hat{\mathbf{R}}^{(1)}, \dots \hat{\mathbf{R}}^{(N_K)}) =\mathrm{SparseMASE}(\mathbf{A}^{(1)}, \dots \mathbf{A}^{(N_K)})$. Subsequently, for $k \in [N_K]$, we obtain $\hat{\mathbf{q}}^{(k)}=\mathrm{vec}(\frac{1}{n_K} \hat{\mathbf{R}}^{(k)})$. We construct a localization graph  with neighbourhood parameter $\lambda_K$ on the points $\left\lbrace \hat{\mathbf{q}}^{(k)}
\right\rbrace_{k=1}^{N^*_K}$ as vertices, and then obtain the isomap embeddings $(\hat{z}_1,\dots \hat{z}_l)$ where $l=6$.
We added the plot obtained from the simulation in \textit{Figure \ref{fig:response_consistency}}.
\newline
\newline
Our next simulation experiment is carried out to support \textit{Corollary \ref{Cor:power_convergence}}.
We take the number of graphs associated with responses to be $s=5$, and set the regression parameters at $\alpha=2.0,\beta=5.0$ and $\sigma_{\epsilon}=0.1$. As before, we define the manifold to be $\mathcal{M}=\psi([0.25,1])$ where $\psi:\mathbb{R} \to \mathbb{R}^D$ is $\psi(t)=(t/a,t/b,t/b,t/a)$. Just as in our previous simulation, we define index $K$ such that the
number of nodes $n_K=16+4(K-1)$,
total number of graphs $N_K=12+(K-1)$, number of graphs for isomap $N^*_K=
\left\lfloor
N_K^{0.85}
\right\rfloor
$ and neighbourhood parameter $\lambda_K=0.95 \times 0.99^{K-1}$ as $K$ varies in the range $\left\lbrace 1,2,...,20 \right\rbrace$. For every $K$, we generate $100$ Monte Carlo samples of a sequence of COSIE graphs and perform the following procedure on every sample. We generate $t_1, \dots t_{N_K} \sim^{iid} \mathrm{Uniform}(0.25,1)$, and for $k \in [s]$, we generate the responses $y_k=\alpha+\beta t_k +\epsilon_k$, where $\epsilon_k \sim^{iid} N(0,\sigma_{\epsilon}^2)$. 
We also compute the predicted responses $\hat{y}_k, k \in [s]$ and the $F$-statistic given by 
$
F^*=(s-2)
        \frac
        {
        \sum_{k=1}^s (\hat{y}_k-\Bar{y})^2
        }
        {
        \sum_{k=1}^s (y_k-\hat{y}_k)^2
        }
$.
For each $t_i$, we form the block connection probability matrix $B(t_i)$ whose diagonal elements are $\frac{t_i}{2}$ and the off-diagonal elements are $\frac{t_i}{5}$. We form the common community membership matrix $\mathbf{Z} \in \mathbb{R}^{n_K \times d}$ by the following rule: the first $\frac{n_K}{2}$ rows are all $(1,0)^T$ and the rest $\frac{n_K}{2}$ rows are all $(0,1)^T$. Thereafter we construct the probability matrices $\mathbf{P}^{(k)}=\mathbf{Z}\mathbf{B}(t_k) \mathbf{Z}^T$, and we sample the adjacency matrices $\mathbf{A}^{(k)}$ such that for all $k \in [N_K]$, and for all $i<j, i,j \in [n_K]$, $\mathbf{A}^{(k)}_{ij} \sim \mathrm{Bernoulli}(\mathbf{P}^{(k)}_{ij})$. Since COSIE model is generalization to multilayer stochastic blockmodel, we have 
$(\mathbf{A}^{(1)}, \dots \mathbf{A}^{(N_K)} \sim \mathrm{COSIE(\mathbf{V},\mathbf{R}^{(1)}, \dots \mathbf{R}^{(N_K)})})$ where $\mathbf{V} \in \mathbb{R}^{n_K \times d}$ has orthonormal columns and every $\mathbf{R}^{(k)}$ is symmetric. After that, we obtain $(\hat{\mathbf{R}}^{(1)},\dots \hat{\mathbf{R}}^{(N_K)})=\mathrm{SparseMASE}(\mathbf{A}^{(1)},\dots \mathbf{A}^{(N_K)},d)$ and then compute $\hat{\mathbf{q}}^{(k)}=\mathrm{vec}(\frac{1}{n_K}\mathbf{R}^{(k)})$ for all $k \in [N_K]$. We construct a localization graph with neighbourhood parameter $\lambda_K$ on the points $\left\lbrace \hat{\mathbf{q}}^{(k)} \right\rbrace_{k=1}^{N_K^*}$, and compute the isomap embeddings $(\hat{z}_1, \dots \hat{z}_s)$. Using a linear regression model on $(y_k,\hat{z}_k)_{k=1}^s$ we 
calculate the predicted responses $\Tilde{y}_k, k \in [s]$ and subsequently
compute the approximate $F$-statistic given by $\hat{F}=
        (s-2)
        \frac
        {
        \sum_{k=1}^s (\Tilde{y}_k-\Bar{y})^2
        }
        {
        \sum_{k=1}^s
        (y_k-\Tilde{y}_k)^2
        }$.
We compute the empirical estimates $\pi^*$ and $\hat{\pi}$ of the powers  of the tests by calculating the proportions (out of $100$ Monte Carlo samples) $F^*$ and $\hat{F}$ exceed a threshold $c_{\Tilde{\alpha}}$, where $c_{\Tilde{\alpha}}$ is the $(1-\Tilde{\alpha})$-th quantile of the $F_{1,s-2}$ distribution and the level of significance is taken $\Tilde{\alpha}=0.05$. The plot is given in \textit{Figure \ref{fig:power_convergence}}. 
\newline
\newline
The following section demonstrates the use of our proposed algorithm on real-world data. 
\section{Analysis of biological learning networks}
\label{Sec:Real_Data}
\begin{figure}[h!]
    \centering
    \includegraphics[scale=1.0]{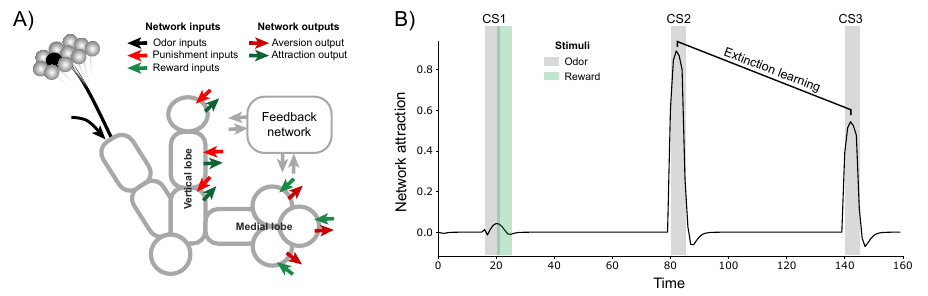}
    \caption{Computational model of the associative learning networks \textit{Drosophila} larvae brains. A) A schematic of the mushroom body (MB), a learning structure in the insect brain composed of the vertical and medial lobes (adapted from \cite{eschbach2020recurrent}). B) Example model output during a extinction learning trial (defined as the ratio of strength of activity in neurons responsible for attraction to strength of activity in neurons responsible for aversion).
    }
    \label{fig:MB_figure}
\end{figure}

\begin{figure}[h!]
    \centering
    \includegraphics[scale=0.85]{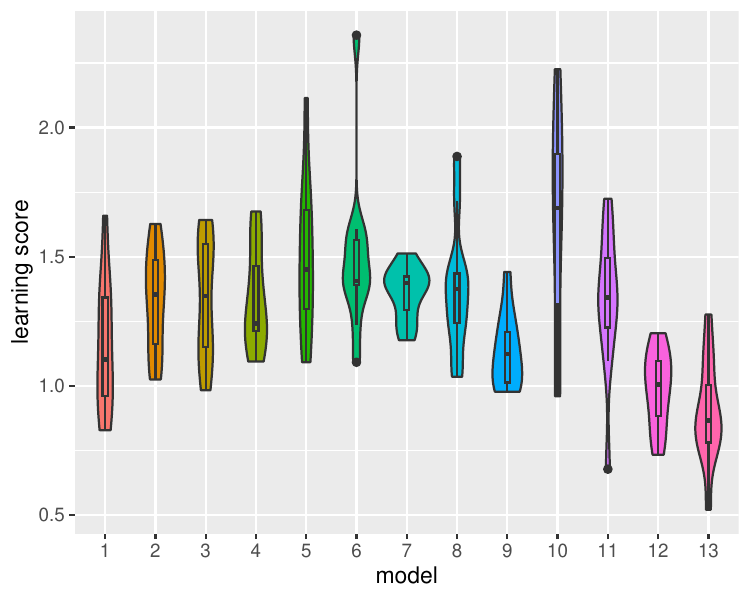}
    \caption{Multiple violin plots of learning scores against corresponding models. Each model of the larval drosophila connectome has $11$ replications, and there are $13$ such models. The violin plots indicate that the learning scores vary significantly across models.}
    \label{fig:model_vs_score_violin}
\end{figure}

\begin{figure}[h!]
    \centering
    \includegraphics[scale=0.90]{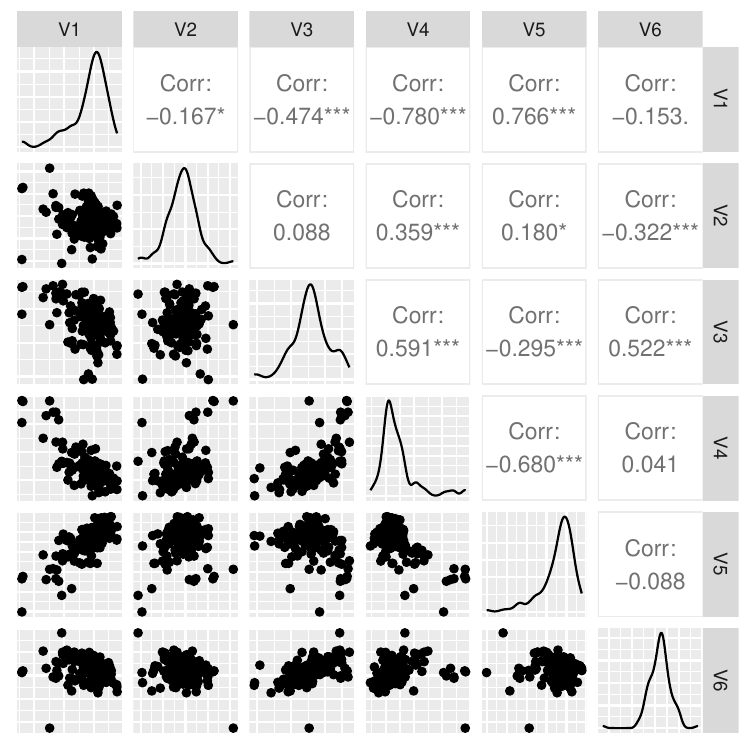}
    \caption{Matrix of scatterplots of (in lower diagonal) and correlation coefficients (in upper diagonal) of all possible pairs in the upper triangle (including diagonal) of the estimated scaled $3 \times 3$ score matrices. The scatterplot indicates the presence of an underlying manifold structure.}
    \label{fig:real_all_matrix}
\end{figure}

\begin{figure}[h!]
    \centering
    \includegraphics[scale=0.65]{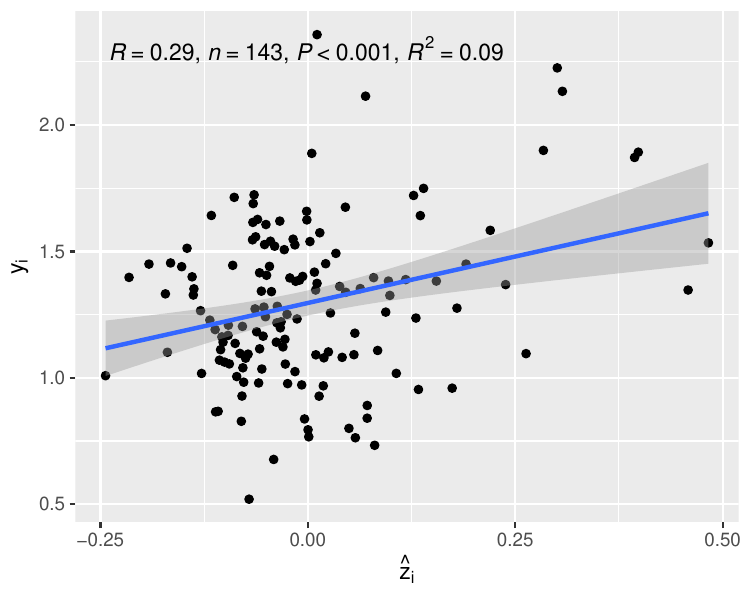}
    \caption{Scatterplot of the learning scores as responses and the one-dimensional isomap embeddings as regressors, along with the fitted linear regression line, when the graphs are selected from the $40$-th positions of the time series. The p-value corresponding to the $F$-test is approximately $0.0004$, suggesting that a simple linear regression model can be used to explain the relationship between the learning scores and the isomap embeddings. The predicted responses are obtained by $\hat{y}_z=1.30+0.74z$.}
    \label{fig:real_response_vs_isomap_embeddings}
\end{figure}

\begin{figure}[h!]
    \centering
    \includegraphics[scale=0.65]{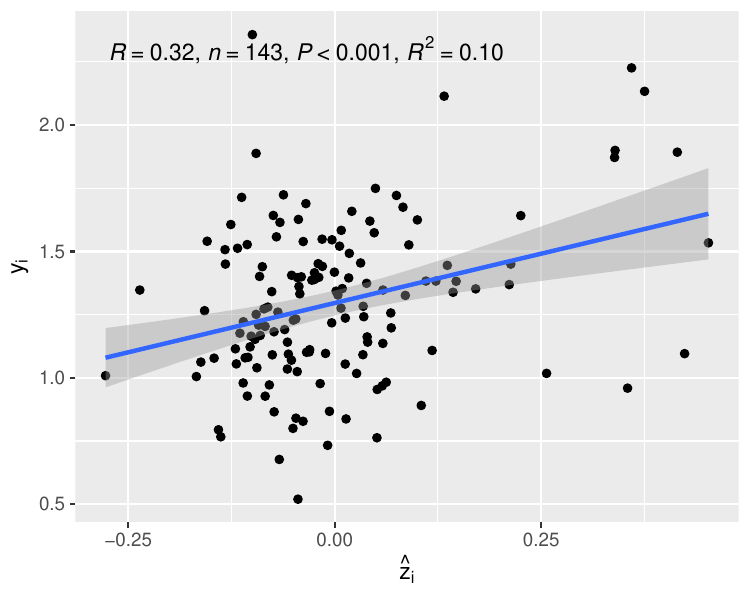}
    \caption{Scatterplot of the learning scores as responses and the one-dimensional isomap embeddings as regressors, along with the fitted linear regression line, when the graphs are selected from the $17$-th position of the time series. The p-value corresponding to the $F$-test is approximately $0.00009$, suggesting that a simple linear regression model can be used to explain the relationship between the learning scores and the isomap embeddings. The predicted responses are obtained by $\hat{y}_z=1.30+0.78z$.}
    \label{fig:real_response_vs_isomap_embeddings_t17}
\end{figure}

\begin{figure}[h!]
    \centering
    \includegraphics[scale=0.65]{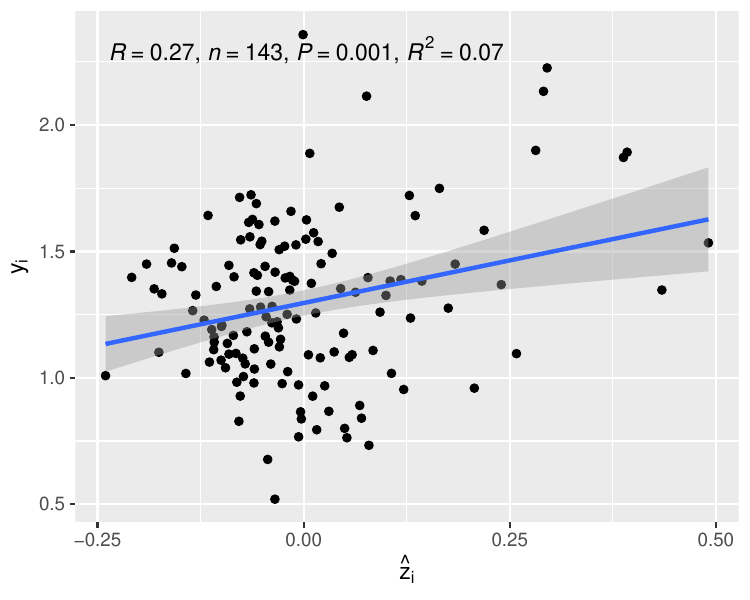}
    \caption{Scatterplot of the learning scores as responses and the one-dimensional isomap embeddings as regressors, along with the fitted linear regression line, when the graphs are selected from the $30$-th position of the time series. The p-value corresponding to the $F$-test is approximately $0.001$, suggesting that a simple linear regression model can be used to explain the relationship between the learning scores and the isomap embeddings. The predicted responses are obtained by $\hat{y}_z=1.30+0.67z$.}
    \label{fig:real_response_vs_isomap_embeddings_t30}
\end{figure}

In this section, we demonstrate the use of our methodology for analysing functional activity in biological learning networks of the \textit{Drosophila} larvae. The complete wiring diagram (or `connectome') of the larval brain was recently completed (\cite{winding2023connectome}), allowing the generation of biologically realistic models of these neural circuits based on known anatomical connectivity \cite{eschbach2020recurrent}). Recent papers have studied how learning networks in this larval brain might operate in the real animal, by training connectome-constrained models to perform associative learning in simulations. In these simulations, a given sequence of stimuli is delivered to the network, that are understood to generate a certain network output in the real animal (e.g. when an odor is paired with pain, the odor becomes less attractive to the animal). 
\newline
\newline
Specifically, we train the network models to perform extinction learning. This is the phenomenon where, after learning an association between a conditioned stimulus (CS; e.g. an odor) and reinforcement (pain or reward), that association is weakened by exposure to the same stimulus in the absence of reinforcement. To do this, we simulate the activity of the network for 160 time points, constituting a time series which corresponds to a single extinction learning trial. 
We perform $143$ such trials corresponding to $11$ replications (from $11$ different randomization seeds) of each of $13$ different models, where each model originates from removal of a single synapse from the parent network (multiple violin plots of learnig scores against models are given in \textit{Figure \ref{fig:model_vs_score_violin}}).
In each trial, at $t=16$, a random odor is delivered to the neurons in the mushroom body (CS1). At $t=20$, either a punishment or a reward is delivered to the mushroom body neurons. All stimuli lasted for $3$ time points. After this pairing of odor with reinforcement, the odor was delivered again at $t=80$ (CS2) and $t=140$ (CS3) (see \textit{Figure \ref{fig:MB_figure}B}).
\newline
\newline
For every extinction learning trial, a learning score is recorded. It is defined as the ratio of network response at the third conditioned stimulus to that at the second conditioned stimulus, where network response at a particular time-point is defined as the ratio of the degree of aversion to that of attraction (for further details about the training of these models, see \cite{eschbach2020recurrent}).
\newline
\newline
We thus obtain $143$ different time series, each consisting of $160$ networks and corresponding to a learning score (where each network has $140$ nodes).
 In our paper, we select one particular graph from each time series , and consider the set of graphs thus selected, associated with the corresponding learning scores. We wish to investigate if the graphs in some manner correspond to points on a low-dimensional manifold in a higher-dimensional ambient space, and if so, find an appropriately simple model to explain the relationship between the learning scores and the pre-images of the points on the manifold.
\newline
\newline
Each time series of graphs has $160$ directed and weighted graphs where each graph consists of $140$ nodes.
From each time series, we select the $40$-th graph and thus form a collection of $143$ weighted and directed graphs, each associated with a response. We then transform the graphs into undirected graphs by ignoring the direction of their edges, and we transform each graph into an unweighted graph by the following rule: if the modulus of the edge weights exceed a particular threshold, then it is stored as one, and otherwise it is stored as zero.
The threshold for censoring the adjacency matrices is chosen to be the $25$-th percentile of the absolute values of the non-zero entries of the weighted adjacency matrix. Upon obtaining the unweighted and binarized form of the adjacency matrices, we apply  \textit{Algorithm \ref{Algo:MASE_sparse}} to get $3 \times 3$ scaled score matrices. 
Since the score matrices are symmetric, the six entries in the upper triangle (including the diagonal) determines the entire matrix.  
We obtain a $6 \times 6$ matrix of scatterplots between all the pairs of components in the upper triangle (including the diagonal) of the estimated scaled score matrices, along with the corresponding correlation coefficients. The plot is given in \textit{Figure \ref{fig:real_all_matrix}}.
\newline
\newline
\newline
\newline
\newline
Suspecting an underlying manifold structure, we obtain $143$ six-dimensional vectors by concatenating 
the entries in the upper triangle of the estimated scaled score matrices, and embed them into one-dimension by using isomap. We then link the responses with the one-dimensional embeddings via a simple linear regression model, and test for the usefulness of the model using $F$-test. The p-value is found to be approximately $0.0004$, thus letting us conclude at $0.05$ level of significance that the simple linear regression model can be used to explain the relationship between the responses and the isomap embeddings. By virtue of \textit{Corollary \ref{Cor:power_convergence}}, this implies that the responses are linked to the pre-images of the score matrices via a simple linear regression model. 
A scatterplot of the responses against the one-dimensional isomap embeddings, along with the fitted regression line, is given in \textit{Figure \ref{fig:real_response_vs_isomap_embeddings}}.
We repeat the abovementioned procedure (previously applied on the collection of graphs at the $40$-th positions of all the time series) for the collection of the networks located at the $17$-th positions of all the time series, and then for the collection of networks at the $30$-th positions of all the time series. 
For both the cases, we conclude (at $0.05$ level of significance) that a simple linear regression model can capture the relationship between the learning scores and the one-dimensional isomap embeddings. The plots are given in \textit{Figure \ref{fig:real_response_vs_isomap_embeddings_t17}} and \textit{Figure \ref{fig:real_response_vs_isomap_embeddings_t30}}.
\newline
\newline
The above demonstrations show how our method can be used to capture the relationship between the learning scores and the collection of networks from $11$ different replications of $13$ different models at some specified time. In all the three abovementioned cases, 
our method suggests that
a simple linear regression model can be used to predict the learning score for a newly obtained graph. 
The following section discusses concisely the overall contribution of this paper, along with certain recommendations in specific scenarios and some possible future extensions. 
\section{Discussion}
\label{Sec:Discussion}
\begin{figure}[h!]
    \centering
    \includegraphics[scale=0.65]{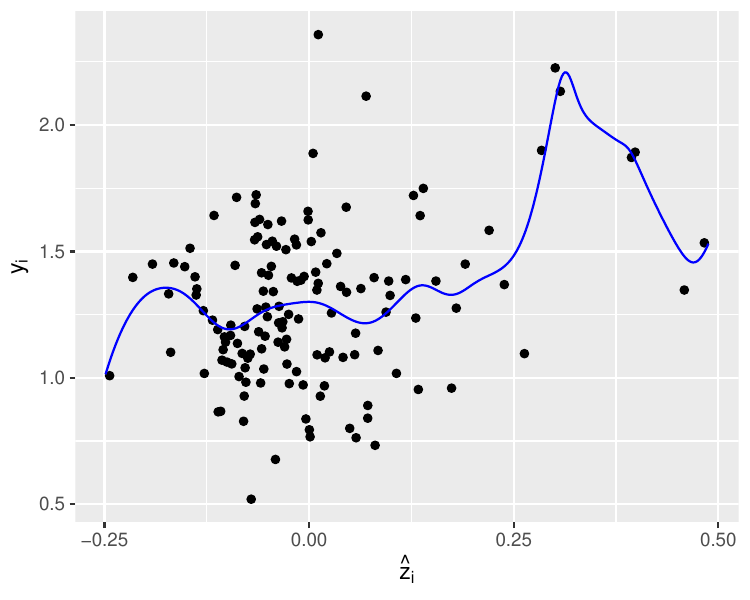}
    \caption{Scatterplot of the learning scores as responses and the one-dimensional isomap embeddings as regressors, along with the fitted nonparametric
    locally linear
    regression line, when the graphs are selected from the $40$-th position of the time series. The bandwidth is $0.03$. The value of R-square is $0.24$.}
\label{fig:real_response_vs_isomap_nonparametric}
\end{figure}

In this article, we propose a method to predict responses corresponding to networks in a semisupervised setting under particular model assumptions. We assume that a large number of networks sampled from the \textit{common subspace independent edge} model (\cite{JMLR:v22:19-558}) are observed and corresponding to only a few amongst them, responses are recorded. Assuming that the networks correspond to points on an unknown one-dimensional manifold in a higher dimensional ambient space, we propose an algorithm exploiting the underlying manifold structure to consistently predict the response at an unlabeled network. 
\newline
\newline
We demonstrate the application of our methodology in real-world data in \textit{Section \ref{Sec:Real_Data}}. A connectome dataset (\cite{winding2023connectome}, \cite{eschbach2020recurrent}) of larval \textit{Drosophila} is considered to demonstrate the use of our algorithm. In a collection of networks associated with responses, we find particular entries of the representative matrices that can be viewed as noisy versions of points on an one-dimensional manifold. By virtue of our results, we conclude that a simple linear  regression model can be used to capture the relationship between the responses and scalar pre-images of the points on the manifold. 
\newline
\newline
The justification for our method rests on the theoretical guarantee of vanishing uniform bound on the regressors. This guarantee can help extend the results to the regime where the responses are linked to scalar pre-images via a nonparametric regression model instead of a simple linear regression. We provide an example from our real data analysis of using a nonparametric regression model to predict the responses. 
We fit a nonparametric (local linear) regression model to the data obtained from selecting the networks at the $40$-th positions of all the time series from our connectome dataset (described in \textit{Section \ref{Sec:Real_Data}}). To be more specific, we select the graph at the $40$-th position from each of the $143$ available time series, estimate the scaled $3 \times 3$ score matrices, vectorize the entries in the upper triangles of $143$ such matrices, apply \textit{isomap} to the $143$ six-dimensional vectors, and fit a nonparametric regression model with the learning scores as responses and the one-dimensional isomap embeddings as regressors. The plot is attached below in \textit{Figure \ref{fig:real_response_vs_isomap_nonparametric}}.
\newline
\newline
While our work establishes asymptotic convergence guarantees
for the proposed algorithm, in real-life one may be constrained to deal with only a small number of graphs, which is why we believe making finite-sample improvements to our algorithm involves an interesting research problem. Moreover, the underlying manifold can have innate dimension higher than one, and investigating that regime and generalizing our work to that extended regime is an open problem. 

\textbf{\large Conflict of interest statement:} The authors declare that there are no conflicts of interests.
\bibliographystyle{elsarticle-num}
\nocite{article}
\nocite{Bernstein00graphapproximations}
\nocite{athreya2017statistical}
\nocite{Montgomery}
\nocite{trosset2021rehabilitating}
\nocite{JMLR:v22:19-558}
\nocite{Holland1983StochasticBF}
\nocite{RubinDelanchy2017ASI}
\nocite{Boccaletti2014TheSA}
\nocite{8889404}
\nocite{Vogelstein2011GraphCU}
\nocite{Young2007RandomDP}
\nocite{winding2023connectome}
\nocite{8215766}
\nocite{jones2020multilayer}
\nocite{borg2005modern}
\nocite{sekhon2021result}
\nocite{agterberg2020nonparametric}
\nocite{eschbach2020recurrent}
\nocite{trosset2024continuous}
\nocite{bhattacharyya2018spectral}
\nocite{Almagro2021DetectingTU}
\nocite{omelchenko2022reducing}
\bibliography{ref}
\newpage

\section{Appendix}
\label{Sec:Appendix}
\subsection{Notations}
\label{Subsec:Notations}
In this paper, we shall denote every vector by a bold lower case letter, for instance $\mathbf{u}$. Any vector $\mathbf{u}$, by default, is a column vector.
We will use bold upper case letters like $\mathbf{H}$ to represent matrices. The $i$-th row of matrix $\mathbf{H}$ will be denoted by $\mathbf{H}_{i*}^T$, and the $j$-th column of the same matrix will be denoted by $\mathbf{H}_{*j}$. The Frobenius norm of a matrix $\mathbf{H}$ will be given by $\left\lVert \mathbf{H} \right\rVert_F$. For a matrix $\mathbf{H} \in \mathbb{R}^{n \times d}$ with columns $\mathbf{h}_1, \dots \mathbf{h}_n$, we denote 
$\mathrm{vec}(\mathbf{H})=[\mathbf{h}_1^T, \dots \mathbf{h}_n^T]^T$, that is, 
$\mathrm{vec}(\mathbf{H}) \in \mathbb{R}^{nd}$ is the vector formed by stacking the columns one below another. 
The $(i,j)$-th element of $\mathbf{H}$ will be denoted as $\mathbf{H}_{ij}$. 
The $N \times N$ identity matrix is denoted as $\mathbf{I}_N$, and $\mathbf{J}_N$ denotes the
$N \times N$ matrix of all ones. The notation $\mathbf{0}_{m,n}$ will be used to denote an $m \times n$ matrix whose each entry is zero.
Random vectors will be denoted by upper case (but not bold-faced) letters like $Z$. 
For any natural number $n$, 
the set $\left\lbrace 1,2, \dots n \right\rbrace$ will be denoted as $[n]$.
The set of all $d \times d$ orthogonal matrices will be represented as $\mathscr{O}(d)$. For any matrix $\mathbf{P} \in \mathbb{R}^m$, 
the eigenvalues of $\mathbf{H}$ will be given by 
$\lambda_1(\mathbf{P}) \geq \dots \geq \lambda_m(\mathbf{P})$. If there is no reason for ambiguity, we will omit the $\mathbf{P}$ and will simply denote the eigenvalues by $\lambda_1 \geq \dots \geq \lambda_m$. The largest and the smallest non-zero eigenvalues of $\mathbf{{P}}$ may also be respectively denoted by $\lambda_{max}(\mathbf{P})$ and $\lambda_{min}(\mathbf{P})$. The maximum row sum of any matrix $\mathbf{H} \in \mathbb{R}^{m \times n}$ will be denoted by $\delta(\mathbf{H})=\max_{i \in [m]}
\sum_{j=1}^n \mathbf{H}_{ij}$.
\newline
\subsection{Proofs}
\label{Subsec:Results_proofs}
\begin{lemma}
\label{Lm:sparsity_estimate_consistency}
    Suppose we observe $n \times n$ adjacency matrices $(\mathbf{A}^{(1)},\dots \mathbf{A}^{(N)}) \sim \mathrm{COSIE}(\mathbf{V};\mathbf{R}^{(1)},\dots \mathbf{R}^{(N)};\rho_n)$. Define
    \begin{equation*}
        \hat{\rho}_n=
        \frac{1}{N {n \choose 2}}
        \sum_{k=1}^N \sum_{i<j}
        \mathbf{A}^{(k)}_{ij}
    \end{equation*}.
Then $|\hat{\rho}_n-\rho_n| \to 0$ as $n \to \infty, N \to \infty$.
\end{lemma}
\textbf{Proof:}
Define for every $k \in [N]$,
$\hat{\rho}^{(k)}_n=
\frac{1}{{N \choose 2}}
\sum_{i<j} \mathbf{A}^{(k)}_{ij}
$.
From \textit{Lemma $5$} of \cite{agterberg2020nonparametric},
we know that for every $k \in [N]$, $\lim_{n \to \infty} (\hat{\rho}^{(k)}_n-\rho_n)=0$. 
Thus, for every fixed $N \in \mathbb{N}$, $\lim_{n \to \infty}
\frac{1}{N} \sum_{k=1}^N (\hat{\rho}^{(k)}_n-\rho_n)=0 \implies 
\lim_{n \to \infty} (\frac{1}{N} \sum_{k=1}^N 
\hat{\rho}^{(k)}_n -\rho_n)=0
$. From \cite{sekhon2021result}, we have $\lim_{N,\to \infty,n \to \infty} (\frac{1}{N} \sum_{k=1}^N \hat{\rho}^{(k)}_n-\rho_n)=0$, that is, $\lim_{N \to \infty, n \to \infty} (\hat{\rho}_n-\rho_n)=0$.
\newline
\newline
\begin{lemma}
\label{Lm:scaled_R_consistency}
    Let $N^* \in \mathbb{N}$ be fixed and suppose we observe $N=N^*+\Tilde{N}$ graphs with adjacency matrices
    $(\mathbf{A}^{(1)},\dots \mathbf{A}^{(N^*)}, \dots \mathbf{A}^{(N)}) \sim \mathrm{COSIE}(\mathbf{V}; \mathbf{R}^{(1)}, \dots \mathbf{R}^{(N^*)}, \dots \mathbf{R}^{(N)};\rho_n)$, where
    $\mathbf{V} \in \mathbb{R}^{n \times d}$ and for all $k \in [N]$, $\mathbf{R}^{(k)} \in \mathbb{R}^{d \times d}$.
    Assume that \textit{Assumptions \ref{Asm:Delocalization}, \ref{Asm:Edge_variance}, \ref{Asm:score_matrix_variance}, \ref{Asm:largest_expected_degree_growth}, \ref{Asm:scaled_R_free_of_graph_size}} hold. Define 
$(\hat{\mathbf{R}}^{(1)}, \dots \hat{\mathbf{R}}^{(N)})=\mathrm{SparseMASE}(\mathbf{A}^{(1)}, \dots \mathbf{A}^{(N)},d)$. As 
$n \to \infty$,
$\Tilde{N} \to \infty$, for all
$k,l \in [N^*]$,
\begin{equation*}
    \left\lVert 
    \hat{\mathbf{Q}}^{(h)}-
    \hat{\mathbf{Q}}^{(k)}
    \right\rVert_F
    -
    \left\lVert 
    \mathbf{Q}^{(h)}-
    \mathbf{Q}^{(k)}
    \right\rVert_F
    \to^P 0,
\end{equation*}
where $\mathbf{Q}^{(k)}=\frac{1}{n} \mathbf{R}^{(k)}$ and $\hat{\mathbf{Q}}^{(k)}=\frac{1}{n} \hat{\mathbf{R}}^{(k)}$ for $k \in [N]$.
\end{lemma}
\textbf{Proof:} 
From \textit{Theorem \ref{Th:MASE_R_asy}}, we know that there exists a sequence of matrices $\mathbf{W} \in \mathcal{O}(d)$, such that
as $n \to \infty$, for all $i,j \in [d]$, 
\begin{equation*}
    \frac{1}{\sigma_{k,i,j}}
    \left(
    \hat{\rho}_n
    \hat{\mathbf{R}}^{(k)}-
    \rho_n
    \mathbf{W}^T \mathbf{R}^{(k)} 
    \mathbf{W}+\mathbf{H}^{(k)}
    \right)_{ij} \to^d N(0,1),
\end{equation*}
where $\mathbb{E}(\left\lVert 
\mathbf{H}^{(k)}
\right\rVert_F)=O(\frac{d}{\sqrt{N}})$ and $\sigma^2_{k,i,j}=O(1)$. Observe that $\lim_{n \to \infty} \frac{\hat{\rho}_n}{\rho_n}=1 
\implies
\lim_{n \to \infty} \frac{\rho_n-\hat{\rho}_n}{\sigma_{k,i,j}}=0$. Hence, as $n \to \infty$, for all $i,j \in [d]$, 
\begin{equation*}
   \frac{1}{\sigma_{k,i,j}}
    \left(
    \hat{\rho}_n
    \hat{\mathbf{R}}^{(k)}-
    \hat{\rho}_n
    \mathbf{W}^T \mathbf{R}^{(k)} 
    \mathbf{W}+\mathbf{H}^{(k)}
    \right)_{ij} \to^d N(0,1). 
\end{equation*}
Recalling that $\rho_n=\Omega(\frac{1}{n})$, observe that 
$\mathbb{E}
\left(
\frac{
\left\lVert 
\mathbf{H}^{(k)}
\right\rVert_F}
{
n \hat{\rho}_n
}
\right) \to 0
$ as $n \to \infty, N \to \infty$.
Note that using the fact that $\lim_{n \to \infty} \frac{\hat{\rho}_n}{\rho_n}=1$ (from \textit{Lemma $5$ of \cite{agterberg2020nonparametric}}), we have $\frac{\sigma^2_{k,i,j}}{n^2 \hat{\rho}_n^2} \to 0$ as $n \to \infty$.
Thus, for every $k \in [N^*]$, as $n \to \infty$ and $N \to \infty$, 
\begin{equation*}
    (\hat{\mathbf{Q}}^{(k)}-
    \mathbf{W}^T
    \mathbf{Q}^{(k)} \mathbf{W}) \to^P \mathbf{0}_{d,d}
    \text{ entrywise}.
\end{equation*}
Hence, we have for all $h,k \in [N^*]$, as $N \to \infty$,
\begin{equation*}
\begin{aligned}
\left(
    (\hat{\mathbf{Q}}^{(h)}-
    \hat{\mathbf{Q}}^{(k)})-
    \mathbf{W}^T (\mathbf{Q}^{(h)}-\mathbf{Q}^{(k)})\mathbf{W}
\right) \to^P \mathbf{0}_{d,d} \text{ entrywise} \\
\implies
\left(
\left\lVert
\hat{\mathbf{Q}}^{(h)}-
\hat{\mathbf{Q}}^{(k)}
\right\rVert_F
-
\left\lVert 
\mathbf{Q}^{(h)}-\mathbf{Q}^{(k)}
\right\rVert_F
\right) \to^P 0.
\end{aligned}
\end{equation*}
\newline
\newline
\textbf{Proposition 1.}
\textit{
Let $N^* \in \mathbb{N}$ be fixed and
    suppose $(\mathbf{A}^{(1)},\dots \mathbf{A}^{(N^*)},\dots \mathbf{A}^{(N)}) \sim \mathrm{COSIE}(\mathbf{V};\mathbf{R}^{(1)},\dots \mathbf{R}^{(N^*)},\dots \mathbf{R}^{(N)};\rho_n)$ where
    $\mathbf{V} \in \mathbb{R}^{n \times d}$ is the common subspace matrix and the matrices $\mathbf{R}^{(k)}$ are the symmetric score matrices, and let $\mathbf{q}^{(k)}=\mathrm{vec}(\frac{1}{n}\mathbf{R}^{(k)})$ for all $k \in [N]$. Denote the multiple adjacency spectral embedding outputs by $(\hat{\mathbf{R}}^{(1)}, \dots \hat{\mathbf{R}}^{(N)})=\mathrm{SparseMASE}(\mathbf{A}^{(1)},\dots \mathbf{A}^{(N)},d)$ and subsequently define $\hat{\mathbf{q}}^{(k)}=\mathrm{vec}(\frac{1}{n}\hat{\mathbf{R}}^{(k)})$ for $k \in [N]$. Assume that $d_{\hat{G},\lambda}(\hat{\mathbf{q}}^{(h)},\hat{\mathbf{q}}^{(k)})$ denotes the shortest path distance between $\hat{\mathbf{q}}^{(h)}$ and $\hat{\mathbf{q}}^{(k)}$ in the localization graph $\hat{G}$ with neighbourhood parameter $\lambda$ constructed on the points $\left\lbrace \hat{\mathbf{q}}^{(k)}\right\rbrace_{k=1}^{N^*}$, and
    $d_{\Tilde{G},\lambda}(\mathbf{q}^{(h)},\mathbf{q}^{(k)})$ denotes the shortest path distance between $\mathbf{q}^{(h)}$ and $\mathbf{q}^{(k)}$ in the localization graph $\Tilde{G}$ with neighbourhood parameter $\lambda$ constructed on the points $\left\lbrace \mathbf{q}^{(k)}\right\rbrace_{k=1}^{N^*}$. Define
    \begin{equation*}
    \hat{\boldsymbol{\Delta}}=
    \left(
    d_{\hat{G},\lambda}
    (\hat{\mathbf{q}}^{(h)},
    \hat{\mathbf{q}}^{(k)})
    \right)_{h,k=1}^l,
    \hspace{0.2cm}
    \Tilde{\boldsymbol{\Delta}}=
    \left(
    d_{\Tilde{G},\lambda}
    (\mathbf{q}^{(h)},\mathbf{q}^{(k)})
    \right)_{h,k=1}^l \in \mathbb{R}^{l \times l}.
\end{equation*}
For any $\lambda>0$, as $n \to \infty$ and $N \to \infty$, 
\begin{equation*}
    \left\lVert 
    \hat{\boldsymbol{\Delta}}-
    \Tilde{\boldsymbol{\Delta}}
    \right\rVert_F \to 0.
\end{equation*}
}
\newline
\textbf{Proof:}
Note that $\left\lVert 
\hat{\mathbf{q}}^{(h)}-
\hat{\mathbf{q}}^{(k)}
\right\rVert=
\left\lVert 
\hat{\mathbf{Q}}^{(h)}-
\hat{\mathbf{Q}}^{(k)}
\right\rVert_F
$ and 
$\left\lVert 
\mathbf{q}^{(h)}-
\mathbf{q}^{(k)}
\right\rVert=
\left\lVert 
\mathbf{Q}^{(h)}-
\mathbf{Q}^{(k)}
\right\rVert_F
$ for all $h,k$.
Thus, for fixed $\lambda>0$, 
$
\left(
\left\lVert \hat{\mathbf{q}}^{(h)}- \hat{\mathbf{q}}^{(k)}
\right\rVert
-
\left\lVert 
\mathbf{q}^{(h)}-
\mathbf{q}^{(k)}
\right\rVert
\right)
\to^P 0$ as $n \to \infty$ and $N \to \infty$. 
Now, consider the $\lambda$-neighbourhood localization graphs, namely $\Tilde{G}$ on the points $\left\lbrace 
\mathbf{q}^{(k)}
\right\rbrace_{k=1}^{N^*}$ and $\hat{G}$ on the points $\left\lbrace
\hat{\mathbf{q}}^{(k)}
\right\rbrace_{k=1}^{N^*}$. 
With overwhelming probability, existence (or non-existence) of an edge between $\mathbf{q}^{(h)}$ and $\mathbf{q}^{(k)}$ in $\Tilde{G}$ will imply existence (or non-existence) of an edge between $\hat{\mathbf{q}}^{(h)}$ and $\hat{\mathbf{q}}^{(k)}$ in $\hat{G}$ for sufficiently large $N$.
Suppose there are total $p_0$ many possible paths between $\mathbf{q}^{(h)}$ and $\mathbf{q}^{(k)}$ in $\Tilde{G}$, and suppose the $p$-th path is along
$\left\lbrace
\mathbf{q}^{(i^p_0)},\mathbf{q}^{(i^p_1)},\dots \mathbf{q}^{(i^p_{v_p})} 
\right\rbrace$
where $i_0=h$ and $i^p_{v_p}=k$, for all $p \in [p_0]$. Without loss of generality assume that $p=1$ corresponds to the shortest path. For sufficiently large $N$, with overwhelming probability, there will exist only $p_0$ many possible paths between $\hat{\mathbf{q}}^{(h)}$ and $\hat{\mathbf{q}}^{(k)}$ in $\hat{G}$. Moreover, for all $p \in [p_0]$,
\begin{equation*}
\left(
\sum_{j=1}^{v_p}
    \left\lVert
    \hat{\mathbf{q}}^{(i^p_j)}-
    \hat{\mathbf{q}}^{(i^p_{j-1})}
    \right\rVert
    -
\sum_{j=1}^{v_p}
    \left\lVert
    \mathbf{q}^{(i^p_j)}-
    \mathbf{q}^{(i^p_{j-1})}
    \right\rVert
\right) \to^P 0
\end{equation*}
as $N \to \infty$. Assuming there are no ties in lengths of paths and denoting $L_1=d_{\Tilde{G},\lambda}(\mathbf{q}^{(h)},\mathbf{q}^{(k)})=\sum_{j=1}^{v_1}
    \left\lVert
    \mathbf{q}^{(i^1_j)}-
    \mathbf{q}^{(i^1_{j-1})}
    \right\rVert$, we can see that $\left(\sum_{j=1}^{v_1}
    \left\lVert
    \hat{\mathbf{q}}^{(i^1_j)}-
    \hat{\mathbf{q}}^{(i^1_{j-1})}
    \right\rVert-L_1\right) \to^P 0$ 
    as $N \to \infty$,
    and for sufficiently large $N$,
    $\sum_{j=1}^{v_p}
    \left\lVert
    \hat{\mathbf{q}}^{(i^p_j)}-
    \hat{\mathbf{q}}^{(i^p_{j-1})}
    \right\rVert>L_1$ with overwhelming probability, when $p \in \left\lbrace 
    2, \dots p_0
    \right\rbrace$. Thus, 
    $\left(
    d_{\hat{G},\lambda}(\hat{\mathbf{q}}^{(h)},\hat{\mathbf{q}}^{(k)})-
    d_{\Tilde{G},\lambda}(\mathbf{q}^{(h)},\mathbf{q}^{(k)})
    \right) \to^P 0
    $ as $N \to \infty$. Since we can choose arbitrary $h, k \in [l]$, we have
    $
    \left\lVert 
    \hat{\boldsymbol{\Delta}}-
    \Tilde{\boldsymbol{\Delta}}
    \right\rVert_F \to 0
    $ as $N \to \infty$.
\newline
\newline
\newline
\textbf{Proposition 2.}
\textit{
Let $l \in \mathbb{N}$ be fixed.
Using the notation from \textit{Proposition \ref{Prop:shortest_path_distance_convergence}},
    suppose 
    \newline
    $(\mathbf{A}^{(1)},\dots \mathbf{A}^{(N^*)},\dots \mathbf{A}^{(N)}) \sim \mathrm{COSIE}(\mathbf{V};\mathbf{R}^{(1)},\dots \mathbf{R}^{(N^*)},\dots \mathbf{R}^{(N)};\rho_n)$. Assume that for all $k$ the points $\mathbf{q}^{(k)}$ lies on an one-dimensional non-self-intersecting compact Riemannian manifold $\mathcal{M}$, and let $d_M(\mathbf{q}^{(h)},\mathbf{q}^{(k)})$ denote the geodesic distance between the points $\mathbf{q}^{(h)}$ and $\mathbf{q}^{(k)}$.
    There exist sequences 
    $\left\lbrace \lambda_K \right\rbrace_{K=1}^{\infty}$ of neighbourhood parameters,
    $\left\lbrace n_K \right\rbrace_{K=1}^{\infty}$ of graph size, $\left\lbrace N_K \right\rbrace_{K=1}^{\infty}$ 
    of total number of graphs
    and 
    $\left\lbrace N^*_K\right\rbrace_{K=1}^{\infty}
    $ of number of graphs for isomap,
    satisfying $\lambda_K \to 0, n_K \to \infty,
    N^*_K \to \infty,
    N_K \to \infty,
    N_K=\omega(N^*_K)
    $ as $K \to \infty$,
    such that when $K \to \infty$,
    \begin{equation*}
        \left\lVert 
        \hat{\boldsymbol{\Delta}}-
        \boldsymbol{\Delta}
        \right\rVert \to 0
    \end{equation*}
    where $\boldsymbol{\Delta}=\left(
    d_M(\mathbf{q}^{(h)},\mathbf{q}^{(k)})
    \right)_{h,k=1}^l$ and
    $
    \hat{\boldsymbol{\Delta}}=
    d_{N^*_K,\lambda_K}
    (\hat{\mathbf{q}}^{(h)},
    \hat{\mathbf{q}}^{(k)})
    $.
}
\newline
\textbf{Proof:} Note that by triangle inequality, we have
$\left\lVert \hat{\boldsymbol{\Delta}} -
\boldsymbol{\Delta}
\right\rVert
\leq 
\left\lVert
\hat{\boldsymbol{\Delta}}-
\Tilde{\boldsymbol{\Delta}}
\right\rVert +
\left\lVert
\Tilde{\boldsymbol{\Delta}}-
\boldsymbol{\Delta}
\right\rVert
$. Observing that $\boldsymbol{\Delta}$, $\hat{\boldsymbol{\Delta}}$ and $\Tilde{\boldsymbol{\Delta}}$ are in general functions of 
total number of graphs $N$, number of graphs for isomap $N^*$, graph size $n$ and neighbourhood parameter $\lambda$, 
denote by $a_{\lambda,N^*,N,n}=
\left\lVert
\hat{\boldsymbol{\Delta}}-
\boldsymbol{\Delta}
\right\rVert
$,
$
b_{\lambda,N^*,N,n}=
\left\lVert
\hat{\boldsymbol{\Delta}}-
\Tilde{\boldsymbol{\Delta}}
\right\rVert
$ and 
$
c_{\lambda,N^*,N,n}=
\left\lVert
\Tilde{\boldsymbol{\Delta}}-
\boldsymbol{\Delta}
\right\rVert
$
when $\hat{\boldsymbol{\Delta}}$ and $\Tilde{\boldsymbol{\Delta}}$ are dissimilarity matrices of shortest path distances in localization graphs constructed on $\left\lbrace \hat{\mathbf{q}}^{(k)} \right\rbrace_{k=1}^{N^*}$ and $\left\lbrace \mathbf{q}^{(k)} \right\rbrace_{k=1}^{N^*}$ respectively, with $\hat{\mathbf{q}}^{(k)}=\mathrm{vec}(\frac{1}{n} \hat{\mathbf{R}}^{(k)})$ being the vectorized forms of scaled multiple adjacency spectral embedding outputs $(\hat{\mathbf{R}}^{(1)},\dots \hat{\mathbf{R}}^{(N)})=\mathrm{SparseMASE}(\mathbf{A}^{(1)},\dots \mathbf{A}^{(N)},d)$, where the adjacency matrices 
$(\mathbf{A}^{(1)},\dots \mathbf{A}^{(N)}) \sim \mathrm{COSIE}(\mathbf{V};\mathbf{R}^{(1)},\dots \mathbf{R}^{(N)})$, $\mathbf{V} \in \mathbb{R}^{n \times d}$ being the common subspace matrix and $\mathbf{R}^{(k)} \in \mathbb{R}^{d \times d}$ for all $k \in [N]$ being the symmetric score matrices. 
Observe that for every fixed $\lambda>0$ and $N^* \in \mathbb{N}$, $\lim_{N \to \infty, n \to \infty} b_{\lambda,N^*,N,n}=0$ from
\textit{Proposition \ref{Prop:shortest_path_distance_convergence}}.
Moreover, observe that by
virtue of
\textit{Assumption \ref{Asm:scaled_R_free_of_graph_size}},
$c_{\lambda,N^*,N,n}$ does not depend on $n$ or $N$, and hence for every
$N \in \mathbb{N}$,
, $n \in \mathbb{N}$,
$c_{\lambda,N^*} \equiv c_{\lambda,N^*,N,n} \to 0$ as
$\lambda \to 0$ and
$N^* \to \infty$, from \textit{Theorem \ref{Th:geodesic_by_shortest_path}}. Define $d_{\lambda,N^*,N,n}=b_{\lambda,N^*,N,n}+c_{\lambda,N^*,N,n}$. For any arbitray $\epsilon>0$, we have 
\newline
(i) for every $\lambda>0$ and $N^* \in \mathbb{N}$, there exists $N^{\prime} \equiv N^{\prime}(\lambda,N^*,\epsilon) \in \mathbb{N}$ such that whenever $N>N^{\prime},n>N^{\prime}$, 
$b_{\lambda,N^*,N,n}<\frac{\epsilon}{2}$.
\newline
(ii) for every tuple $(N,n) \in \mathbb{N}^2$, there exists $\lambda^{\prime\prime} \equiv \lambda^{\prime\prime}(\epsilon) >0$ and $N^{\prime\prime} \equiv N^{\prime\prime}(\epsilon) \in \mathbb{N}$ such that whenever $\lambda<\lambda^{\prime\prime}$ and $N^*>N^{\prime\prime}$, $c_{\lambda,N^*,N,n}<\frac{\epsilon}{2}$ (since $c_{\lambda,N^*,N,n}$ does not depend on $n$ or $N$, nor does $N^{\prime\prime}$).
\newline
Thus, whenever
$\lambda<\lambda^{\prime\prime}(\epsilon)$,
$N^*>N^{\prime\prime}(\epsilon)$ and $N,n>N^{\prime}(N^*,\epsilon)$, $d_{\lambda,N^*,N,n}<\epsilon$. This means there exist sequences $\left\lbrace N_K \right\rbrace_{K=1}^{\infty}$, $\left\lbrace N^*_K \right\rbrace_{K=1}^{\infty}$, $\left\lbrace n_K \right\rbrace_{K=1}^{\infty}$ and $\left\lbrace \lambda_K \right\rbrace_{K=1}^{\infty}$
satisfying $N_K \to \infty, N^*_K \to \infty,n_K \to \infty, \lambda_K \to 0$ as $K \to \infty$, such that
$\lim_{K \to \infty} d_{\lambda_K,N^*_K,N_K,n_K}=0$. Noting that $0 \leq a_{\lambda_K,N^*_K,N_K,n_K} \leq d_{\lambda_K,N^*_K,N_K,n_K}$, we can say $\lim_{K \to \infty} a_{\lambda_K,N^*_K,N_K,n_K}=0$.
\newline
\newline
\newline
\textbf{Theorem 4.}
\textit{
Suppose we have $N$ graphs with adjacency matrices 
$(\mathbf{A}^{(1)}, \dots \mathbf{A}^{(N)}) \sim \mathrm{COSIE}(\mathbf{V};\mathbf{R}^{(1)}, \dots \mathbf{R}^{(N)},\rho_n)$. 
Define $\mathbf{q}^{(k)}=\mathrm{vec}(\mathbf{Q}^{(k)})$ where $\mathbf{Q}^{(k)}=\frac{1}{n} \mathbf{R}^{(k)}$, and assume for all $k$, $\mathbf{q}^{(k)}=\psi(t_k)$ lies on the one-dimensional manifold $\mathcal{M}=\psi([0,L])$. 
Let $s \ll N$ be fixed and responses $y_1, \dots y_s$ are recorded at the first $s$ graphs, and assume the following regression model holds:
\begin{equation*}
    y_k=\alpha+\beta t_k + \epsilon_k 
\end{equation*}
where $\epsilon_k \sim^{iid} N(0,\sigma^2_{\epsilon})$ for all $k \in [s]$.
Suppose \textit{Assumptions \ref{Asm:Delocalization}, \ref{Asm:Edge_variance},
\ref{Asm:score_matrix_variance},
\ref{Asm:largest_expected_degree_growth},
\ref{Asm:scaled_R_free_of_graph_size}
}
hold. Denote by $\hat{\mathbf{q}}^{(k)}=\mathrm{vec}(\hat{\mathbf{Q}}^{(k)})$, where $\hat{\mathbf{Q}}^{(k)}=\frac{1}{n}\hat{\mathbf{R}}^{(k)}$. There exists a sequence $N_K$ of total number of graphs,
$N^*_K=o(N_K)$ of number of graphs for isomap,
and $\lambda_K$ of neighbourhood parameters, for which $N^*_K \to \infty$, $N_K \to \infty$ and $\lambda_K \to 0$ as $K \to \infty$, such that 
for a fixed $l \in \mathbb{N}$, the predicted response $\Tilde{y}_r=\mathrm{PredGraphResp}(\left\lbrace\mathbf{A}^{(k)}\right\rbrace_{k=1}^{N_K}, 
\left\lbrace y_k \right\rbrace_{k=1}^s,d,
\lambda_K,N^*_K,l,r)$ (see \textit{Algorithm \ref{Algo:predict_response}})
will satisfy: for every $r \leq l$, as $K \to \infty$,
\begin{equation*}
|\Tilde{y}_r-\hat{y}_r| \to^P 0
\end{equation*}
where $\hat{y}_r$ is the predicted response for the $r$-th network based on the true regressors $t_k$.
}
\newline
\textbf{Proof:}
Note that the predicted response 
$\Tilde{y}_r=\mathrm{PredGraphResp}(\left\lbrace 
\mathbf{A}^{(k)}
\right\rbrace_{k=1}^{N_K},
\left\lbrace 
y_k
\right\rbrace_{k=1}^s,
N^*_K,\lambda_K,
d,l,r
)$ is the predicted response corresponding to $\hat{z}_r$ when the bivariate data is $(y_k,\hat{z}_k)_{k=1}^s$, and recall that $\hat{y}_r$ is the predicted response corresponding to $t_r$ when the bivariate training set is $(y_k,t_k)_{k=1}^s$. 
From \textit{Proposition \ref{Prop:MASE_output_shortest_path_to_true_geodesics}}, we can see that $\left\lVert
\hat{\boldsymbol{\Delta}}-
\boldsymbol{\Delta}
\right\rVert \to 0$ as $K \to \infty$. Using \textit{Theorem \ref{Th:dissimilarity_continuity}}, we can obtain embeddings $(\hat{z}_1, \dots \hat{z}_l)=\mathrm{ISOMAP}(
\left\lbrace
\hat{\mathbf{q}}^{(k)}
\right\rbrace_{k=1}^{N^*_K}, \lambda_K,l)
$ such that for every $h,k \in [l]$, 
\begin{equation*}
    |\hat{z}_h-\hat{z}_k|-
    |t_h-t_k| \to 0.
\end{equation*}
Thus, as $K \to \infty$, the isomap embeddings $\hat{z}_k$ are getting arbitrarily closer to some affine transformation on the true regressors $t_k$, and in the setting of simple linear regression the predicted response value remains invariant to affine transformations on regressors. Thus, as $K \to \infty$, 
$|\Tilde{y}_r-\hat{y}_r| \to 0$.
\newline
\newline
\newline
\textbf{Corollary 1:}
\textit{
In the setting of \textit{Theorem \ref{Th:isomap_consistency}}, suppose we are to conduct the test $H_0:\beta=0$ against $H_1:\beta \neq 0$ at level of siginificance $\Tilde{\alpha}$. Define the following test statistics:
    \begin{equation*}
        F^*=(s-2)
        \frac
        {
        \sum_{k=1}^s (\hat{y}_k-\Bar{y})^2
        }
        {
        \sum_{k=1}^s (y_k-\hat{y}_k)^2
        }, \hspace{0.5cm}
        \hat{F}=
        (s-2)
        \frac
        {
        \sum_{k=1}^s (\Tilde{y}_k-\Bar{y})^2
        }
        {
        \sum_{k=1}^s
        (y_k-\Tilde{y}_k)^2
        }.
    \end{equation*}
    Suppose $\pi^*$ is the power of the test done by the rule: reject $H_0$ if $F^*>c_{\Tilde{\alpha}}$, and let $\hat{\pi}$ be the power of the test done by the rule: reject $H_0$ if 
    $\hat{F}>c_{\Tilde{\alpha}}$. Then, for every $(\alpha ,\beta)$,
    $|\hat{\pi}-\pi^*| \to 0$ as $K \to \infty$.
}
\newline
\newline
\textbf{Proof:} We know, from \textit{Theorem \ref{Th:isomap_consistency}} that for any $(\alpha,\beta) \in \mathbb{R}^2$, for all $r \in [s]$, 
$|\Tilde{y}_r-\hat{y}_r| \to^P 0$ as $K \to \infty$. Hence, we must have for any $(\alpha,\beta) \in \mathbb{R}^2$, $|\hat{F}-F^*| \to^P 0$ as $K \to \infty$, and consequently for all $(\alpha,\beta) \in \mathbb{R}^2$ and for all $\Tilde{\alpha} \in (0,1)$, we have
$|\hat{\pi}-\pi^*|=\left|\mathbb{P}_{\alpha,\beta}[\hat{F}>c_{\Tilde{\alpha}}]-\mathbb{P}_{\alpha,\beta}[F^*>c_{\Tilde{\alpha}}]\right| \to 0$ as $K \to \infty$.


\addtolength{\textheight}{.5in}%





\end{document}